\documentclass{aa}
\usepackage{psfig,graphicx,supertabular,amsmath}

\begin{document}
\newcommand{\beq}{\begin{equation}}
\newcommand{\eeq}{\end{equation}}
\newcommand{\beqar}{\begin{eqnarray}}
\newcommand{\eeqar}{\end{eqnarray}}
\newcommand{\excs}{\extracolsep{\fill}}

\thesaurus{11(02.18.5;03.09.1;03.20.4;11.01.2;11.19.1)}
\title{Search for optical microvariability in a large sample of Seyfert I
  galaxies
\thanks{Based on observations taken at the Cananea and San Pedro del M\'artir
  observatories in Mexico}} 
\author{P.O. Petrucci \inst{1}, A. Chelli \inst{1}, G. Henri \inst{1},
  I. Cruz-Gonz\'alez \inst{2}, L. Salas \inst{3}, R. Mujica \inst{4}}
\offprints{P.O. Petrucci}
\mail{Petrucci@obs.ujf-grenoble.fr}
\institute{Laboratoire d'Astrophysique, Observatoire de Grenoble,B.P 53X,
F38041 Grenoble Cedex, France\\
\and 
Instituto de Astronomia, UNAM, Ap. 70-264, C.U., 04510 M\'exico, D.F., M\'exico\\
\and
Instituto de Astronomia, UNAM, Ap. 877, 22830 Ensenada, M\'exico\\
\and
Instituto Nacional de Astrofisica, Optica y Electronica, Ap. 51 y 216
Z.P., 72000 Puebla, M\'exico}
\date{Received ??/ accepted ??}
\maketitle
\markboth{P.O. Petrucci et al.: Search for optical microvariability in
  Seyfert galaxies}{}

\begin{abstract}
  We present results of an optical (I band) monitoring of a sample of 22
  Seyfert I galaxies. We aimed to detect microvariability with time
  resolution from $\simeq$ 6 minutes down to 30 seconds for the most
  luminous one. It is the largest survey ever done in the search of rapid
  optical variations in Seyfert galaxies. We used differential photometry
  and a new method of analysis between galaxy and comparison stars light
  curves in order to minimize the influence of the intrinsic
  variabilities of the latter. We thus obtain precision on standard
  deviation measurements less than 1\% and generally of the order of
  0.5\%. We obtain no clear detection of microvariability in any of
  these objects. In the hypothesis where optical microvariability could
  be due to synchrotron emission of a non thermal electrons population,
  we discuss the physical constraints imposed by these results.

\end{abstract}

\section{Introduction}
If optical microvariability for radio-loud AGNs is now well established
(Carini et al. \cite{Car91}, Miller et al. \cite{Mil92}, Wagner et al.
\cite{Wag92}, Doroshenko et al. \cite{Dor92}), any search for intra-night
optical variability in radio-quiet QSOs (Gopal-Krishna et al
\cite{Gop93}, \cite{Gop95}, Jang et al. \cite{Jan97}, Rabbette et al.
\cite{Rab98}) or Seyfert galaxies in the past ten years does not report
clear evidence of such phenomenon. In fact, much of the controversy may
be due to the transient characteristics of microvariability, precluding a
clear confirmation of any reported variations. This becomes a critical
problem in the case of Seyfert galaxies, since very few observations of
such objects were performed to search for optical flickering. To our
knowledge, the first one was done by Lawrence et al.  (\cite{Law81}).
They observed NGC 4151 on a range of time-scales from 10 seconds to 1
week, but detected no variation to better than $0.05$ mag in bands V and
K. Five years later, systematic photoelectric $UBV$ observations of rapid
variability in AGNs were begun at the Crimean Laboratory in 1986 and have
been carried out for the Seyfert galaxies NGC 4151, NGC 7469, NGC 3516
and NGC 5548 (Lyuty\u{i} et al. \cite{Lyu89}, Aslanov et al.
\cite{Asl89} and Lyuty\u{i} \& Doroshenko \cite{Lyu93} respectively).
Optical microvariabilities were detected in each of these objects: NGC
4151, NGC 7469 and NGC 3516 showed amplitude of microvariability up to
10\% over 15-20 min and about 5\% for NGC 5548 with a shorter time-scale,
i.e.  $\sim$ 10-15 min. But for each cases, the phenomenon is not
continuous and periods without any rapid variabilities are also observed
all along each run. It seemed thus that onset and disappearance of
microvariability follow a random process as observed in radio-loud AGNs
(Carini \cite{Car90}).  A number of other investigators attempted to
detect or confirm the presence of these rapid variations. For example,
the case of NGC 7469 was confirmed by Dultzin-Hacyan et al.
(\cite{Dul92}) althougth no variations were
obtained during another run (Dultzin-Hacyan et al.  \cite{Dul93}). \\
Yet, particulary important results come from simultaneous multifrequency
observations which can put strong constraints on the spatial distribution
of the emitting regions and indicate whether the same radiative process
dominates at different frequencies. The only such search for Seyfert
galaxies is the simultaneous optical-infrared-X-ray study of NGC 4051 by
Done et al. (\cite{Don90}). They report that, on time-scales of tens of
minutes, the flux remained constant within 1\% and 5\% in optical and
infrared, respectively, while the X-ray flux continually flickered by up
to a factor 2. Another survey of this galaxy was done by Hunt et al.
(\cite{Hun92}), but only in the K band, and confirm the upper limit on
nuclear variability of about 2\%.  Done et al. deduced from their results
that the IR/optical source must be at least an order of magnitude larger
than, or completely separate from, the X-ray source.\\
It appears from these results that, in a general manner, the study of
microvariability in Seyfert galaxies is not sufficiently complete to
clearly conclude if optical flickering is (or is not) a common
characteristics for this class of AGNs. This contrasts with the more
complete works done recently with QSOs. First, Jang et al.
(\cite{Jan97}) report, on a selected sample of radio-quiet and radio-loud
QSOs, an apparent contrast in microvariations between the two class of
quasars, 20\% of the radio-quiet objects showing evidence of flickering
against 85\% for radio-loud. Next, Rabbette et al. (\cite{Rab98}) have
just published a search for rapid optical variability (on time-scales of
few minutes) in a large sample of 23 radio-quiet quasars. They report no
detection, with a precision of few percents, of any significant rapid
variability for any of the sources observed.

Presently, no clear explanations of microvariability are approved
unanimously. Unlike radio-loud AGNs where flikering could be due to the
presence of shock inside a relativistic jet (Qian et al. \cite{Qia91},
Gopal-Krishna \& Wiita \cite{Gop92}), no such conclusion can be drawn up
to now for radio-quiet objects such as Seyfert galaxies, since their high
energy spectrum is apparently cut-off above a few hundred keV (Jourdain
et al. \cite{Jou92a}; Maisack et al. \cite{Mais93}; Dermer \& Gehrels
\cite{DermGehr95}). Thus, some models supposed that microvariability
could be due to disturbances (like flares or hot spots) in the accretion
disk surrounding the central engine (Wiita et al \cite{Wii91},
\cite{Wii92}, Chakrabarti \& Wiita \cite{Cha93}, Mangalam \& Wiita
\cite{Man93}). But some results of recent observations do not provide
strong support for such models (Jang \& Miller \cite{Jan95}).  In the
case of Seyfert galaxies, the origin of microvariability could be
associated with the high energy process giving birth to the hard X-ray
spectra (up to few hundred of keV) observed in these galaxies. The source
of the high energy emission is still uncertain: it could be produced
through the comptonization of low energy photons by a thermal, mildly
relativistic plasma $(kT \leq m_ec^2)$ (Haardt \& Maraschi \cite{Haa91})
or by Inverse Compton process from a non-thermal, highly relativistic $(E
\gg m_ec^2)$ particle distribution (possibly made of electron-positron)
(e.g.  Zdziarski et al. \cite{Zdz94}, Henri \& Petrucci \cite{Hen97}). As
there is probably some magnetic field to accelerate and confine the
particles, synchrotron emission is expected to be produced in the latter
case, but not in the former. Purely thermal emission in the optical range
is likely produced in too broad a region to produce intra-day
variability. On the opposite, synchrotron emission should be correlated
to X-ray emission, and thus be also rapidly variable. Therefore a
positive detection of rapid (intra-day) visible-IR variability would
strongly favour non-thermal models. Conversely, non-detection would bring
very valuable upper limits on the intrinsic properties of the local
environment of the emission region, giving strong constraints on the
various models of non thermal emission (Celotti et al. \cite{Cel91}).
Besides, if it exists, the synchrotron emission is diluted by the stellar
contribution and probably by the thermal continuum possibly emitted by an
accretion disk and by dust. The dust emission peaks in the IR range and
fall down near 1 $\mu$m due to dust sublimation, while the disk emission,
supposed to give rise to the Blue Bump, peaks in the UV range.  Hence the
most favourable wavelength domain to detect variable synchrotron emission
would be around 1 $\mu$m.

We present here the results of two observational campaigns of a sample of
22 Seyfert 1 galaxies in the I band at $0.9\mu m$ (and simultaneously in
the J band at $1.25\mu m$ for 3 of them), at the observatories of Cananea
and San Pedro M\'artir in Mexico.  We aim to detect rapid optical
variabilities by differential photometry between the galaxies and the
comparisons stars in the CCD field of view. We have developped a new
method of analysis which minimize the influence of the intrinsic
variabilities of the comparisons stars.  In Section 2 we report on the
sample and the observations. The data analysis method is explained in
Section 3. We present the results for each galaxy, in Section 4,
developing the cases of the more interesting ones.  We will finally
discuss the theoretical constraints imposed by these outcomes in Section
5 before concluding.

\section{Observations}
\begin{table*}
\caption[]{List of observed galaxies. Coordinates and magnitudes are
  taking from V\'eron-Cetty \& V\'eron (\cite{Ver89}) \label{listgal}}
\begin{flushleft}
\begin{tabular}{lllclccclc}
\noalign{\smallskip}
\hline
\noalign{\smallskip}
 Name & RA (1950) & Dec (1950) & z & $m_V$ & Julian day & Exposure time
 & Total duration & Place & Filter \\
 & & & & & & & of the run & & \\
 & & & & &2450000+ & (seconds) & (hours) & & \\
\noalign{\smallskip}
\hline
\noalign{\smallskip}
 Mkn 543         & 23 59 52.9   & +03 04 26    & 0.026   & 14.7 & 423.568
 & 150 & 3.5 & Cananea   & I \\
 Mkn 335         & 00 03 45.1   & +19 55 27    & 0.025   & 13.9 & 420.615
 & 60  & 2.8 & Cananea   & I \\
 Mkn 359         & 01 24 50.1   & +18 55 07    & 0.017   & 14.2 & 419.630
 & 120 & 3.5 & Cananea   & I \\
 Mkn 590         & 02 12 00.5   & -00 59 57    & 0.027   & 13.8 & 421.589
 & 90  & 3.9 & Cananea   & I \\
 Mkn 1044        & 02 27 38.2   & -09 13 11    & 0.016   & 14.3 & 424.578
 & 100 & 4.2 & Cananea   & I \\
 NGC 1019        & 02 35 52.33  & +01 41 32.1  & 0.024   & 14.9 & 425.581
 & 150 & 2.5 & Cananea   & I \\
 Mkn 372         & 02 46 30.9   & +19 05 54    & 0.031   & 14.8 & 422.598
 & 200 & 4.5 & Cananea   & I \\
 IRAS 04448-0513 & 04 44 52.2   & -05 13 33    & 0.044   & 14.6 & 425.752
 & 200 & 3.3 & Cananea   & I \\
 1H 0510+031     & 05 10 03.0   & +03 08 13    & 0.016   & 14.8 & 423.731
 & 260 & 3.6 & Cananea   & I \\
 ARK 120         & 05 13 38.0   & -00 12 17    & 0.033   & 13.9 & 419.834
 & 60  & 3.6 & Cananea   & I \\
 MCG+08-11-11    & 05 51 09.60  & +46 25 50.9  & 0.020   & 14.6 & 420.809
 & 90  & 4 & Cananea   & I \\
 Mkn 376         & 07 10 36.13  & +45 47 06.3  & 0.056   & 14.6 & 424.792
 & 150 & 3.5 & Cananea   & I \\
 Mkn 9           & 07 32 42.4   & +58 52 56    & 0.039   & 14.4 & 422.817
 & 200 & 4.7 & Cananea   & I \\
 PG 0844+349     & 08 44 33.93  & +34 56 08.6  & 0.064   & 14.  & 421.810
 & 120 & 3.3 & Cananea   & I \\
 NGC 4051        & 12 00 36.3   & +44 48 34    & 0.002   & 12.9 & 423.960
 & 40  & 0.5 & Cananea   & I \\
                 &              &              &         &      & 425.926
 & 60  & 1 & Cananea   & I \\
 NGC 4151        & 12 08 01.055 & +39 41 01.82 & 0.003   & 11.8 & 425.990
 & 8   & 0.3 & Cananea   & I \\
 Mkn 1383        & 14 26 33.7   & +01 30 27    & 0.086   & 14.9 & 211.803
 & 120 & 2 & San Pedro & I \\
 Mkn 684         & 14 28 53.1   & +28 30 29    & 0.046   & 14.7 & 212.810
 & 300 & 2.5 & San Pedro & I \\
 Mkn 478         & 14 40 04.59  & +35 39 07.6  & 0.077   & 14.6 & 216.842
 & 210 & 2.3 & San Pedro & I, J \\
 Mkn 1392        & 15 03 25.9   & +03 53 59    & 0.036   & 14.3 & 214.783
 & 300 & 3.3 & San Pedro & I, J \\
 Mkn 1098        & 15 27 37.9   & +30 39 23    & 0.035   & 14.9 & 215.788
 & 300 & 3.3 & San Pedro & I, J \\
 IRAS 15438+2715 & 15 43 52.6   & +27 15 49    & 0.031   & 14.6 & 213.772
 & 300 & 3.3 & San Pedro & I \\
\noalign{\smallskip}
\hline
\end{tabular}
\end{flushleft}
\end{table*}

\subsection{The sample}
The observed galaxies are listed in table \ref{listgal}, along with their
1950 coordinates, redshift $z$, apparent V magnitude, date and place of
observation. These objects have been selected from V\'eron-Cetty \&
V\'eron (\cite{Ver89}) to fulfil the following criteria: a) $-10^\circ
\leq \delta(1950)$, b) $m_v \leq 15$ to reach photometric signal to noise
of about few thousands in a few minutes of integration, c) size $\leq 1$
arcmin to be limited by photon noise and not by readout noise, d) there
must be 3 or more possible comparison sources in the arcmins CCD field,
thus allowing us to identify and discount any of the comparison stars
that are themselves variable on short time-scales. Some of these objects
have at least 2 stars closer than 1 arcmin and were thus suitable for the
infrared camera CAMILA of San Pedro; they were observed simultaneously in
the visible and the IR.  Some objects, like NGC 4051, NGC 4151 and
MCG+08-11-11, did not fulfill all these criteria particularly the c) one
since the size of these galaxies was about the third of the field of
view. But, firstly they are well known objects, already observed for
search of microvariability for two of them (Lyutyi et al. \cite{Lyu89},
for NGC 4151; Done et al.  \cite{Don90}, for NGC 4051) and so interesting
to study.  Secondly, due to their proximity, their flux were high enough
to be rapidly limited by photon noise and not by readout noise in a few
seconds integration time. We could not satisfy criterion d) for NGC 4151
as well, and only one comparison star was in the CCD field. Consequently,
we have treated this galaxy differently (see Section \ref{n4151}).  One
or two photometric standard stars, selected from Landolt (\cite{Lan92})
were also observed before and after each galaxy run to estimate the mean
brightness of the object, needed to deduce physical constraints (see
Section \ref{physicalconstraint}).

\subsection{The observational campaigns}
We have carried out two campaigns of observations in Mexico. The first
one was done in the I band during 7 nights (7-13 May 1996) at the 1.5 m
telescope of the Observatorio Astron\'{o}mico Nacional at San Pedro
M\'artir (Baja California). We used a $1024\times 1024$ Tektronix CCD with
6 electrons readout noise and $4\times 4\ \mbox{arcmin}^2$ field of view.
Simultaneous observations in the J band were performed at the 2.1m
telescope during the 3 last nights. We used the CAMILA $256\times 256$
infrared camera with 40 electrons readout noise and $2\times 2\ 
\mbox{arcmin}^2$ field of view (Cruz-Gonz\`{a}lez et al.  \cite{Cru93}).
The second campaign was done at the 2.1m telescope of the Guillermo Haro
Observatory in Cananea (Sonora) during 8 nights (1-9 December 1996) in
the I band. Only a useful $400\times 600$ pixels part of a $1024\times
1024$ CCD Tektronix, with 8 electrons readout noise and $6\times 10\
\mbox{arcmin}^2$ equivalent field of view, was read.
Galaxies with several comparison stars with comparable brightness in the
field of view are observed as a priority.  The exposure time was chosen
to use the CCD at about half of its dynamic in order to prevent
saturation due to rapid changes of seeing. The acquisition program was
automated to take an image with a period equal to the exposure time plus
the backup time.

\section{Data analysis}
\label{dataanalysis}
\subsection{Fluxes measurement}
\begin{figure}[tbp]
\includegraphics[angle=-90,width=\columnwidth,origin=br]{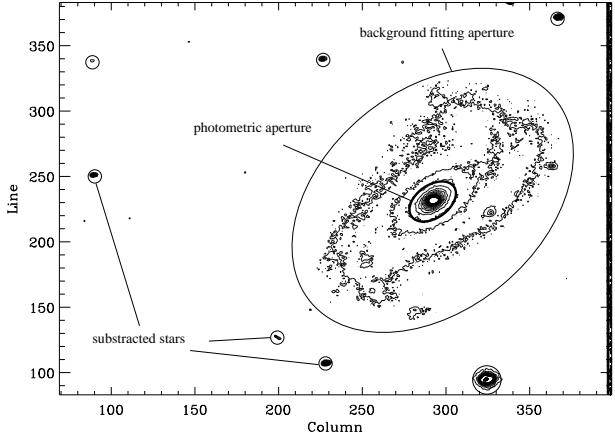}
  \caption[]{Different apertures used to measure the central flux of a
    galaxy (see text). We show the case of NGC 4051, the biggest galaxy
    of our sample.}\label{apertures}
\end{figure}

The individual CCD frames are reduced using standard IRAF software
procedures by substracting the bias frame and by flat-fielding using the
median sky exposures. We choose at least three comparison
stars with about the same brightness than the galaxy in the CCD frame.
Faint sources in their neighbourhood and in the vicinity of the galaxy
are substracted and replaced by the median value measured in annuli
around. Then we use circular apertures to measure the fluxes of the
comparison stars. For galaxies, we can use circular or elliptic apertures
depending on the size and form of the galaxy. In fact, for large galaxies
, like NGC 4051 or NGC 4151, we used two apertures: the first one to fit
the background of the image at the galaxy position (the background
fitting aperture), the second one (the photometric aperture), smaller, to
measure the flux of the central nucleus (see Fig.(\ref{apertures})). In
the more general case, for starlike galaxies, these two apertures are the
same and are circular. \\
In order to fit the sky background in each aperture, we extract a
subimage centred on each object, the size of this subimage being four
times the radius of the background fitting aperture. We fit this subimage
line by line and column by column with a 3 degrees polynomial, using only
points outside the aperture. We take the average of the line by line and
column by column fits to estimate the background flux. This flux is
substracted at the total flux measures within the photometric aperture to
obtain the intrinsic flux of the stars or of the galaxy. We repeat the
treatment for each image of the run, which are recentered, if necessary,
towards a reference image in order to compensate the telescope drifts.

\subsection{Treatment and light curves achievment}
\label{treatsection}
\subsubsection{General case}
Our treatment rests on the small probability that two stars of a given
image vary intrinsically by the same amount from their average behaviour.
If it is the case, the variation is supposed to be due to an extrinsic
perturbation like scintillation, seeing, or atmospheric extinction and
all objects in the field of view are affected in the same way by this
perturbation. It ensues from this that, in this image {\em the two stars
  can play the role of standard stars}.  Actually, due to the different
electronic and statistic noises, we can never detect stars varying
exactly in the same manner. We used thus a minimizing method where the
function to minimize, for a number $n_s$ of comparison stars in the CCD
field, can be expressed as follows (we minimize with respect to the
variable N which plays the role of a normalized flux):
\begin{equation}
\label{minimfunc}
S_{i}(N)=\sum_{j=1}^{n_s}P^j_i(N)
\end{equation}
where
\begin{equation}
\label{Pj}
P^j_i=\prod_{k\neq j}\frac{(\bar{x}_{i,k}-N)^2}{\bar{\sigma}_{i,k}^2}
\end{equation}
In Eq.(\ref{Pj}), $\bar{x}_{i,k}$ and $\bar{\sigma}_{i,k}$ are
respectively the {\em relative} flux (i.e. normalized to the average flux
of the star $k$ on all the images of the run) and the corresponding
relative noise of the comparison star $k$ in the image $i$. The noise
includes the photon and the read-out noises, and is usually dominated by
the former.

\subsubsection{Differences from standard $\chi^2$ reduction}
\begin{figure}[tbp]
\includegraphics[angle=0,width=\columnwidth,origin=br]{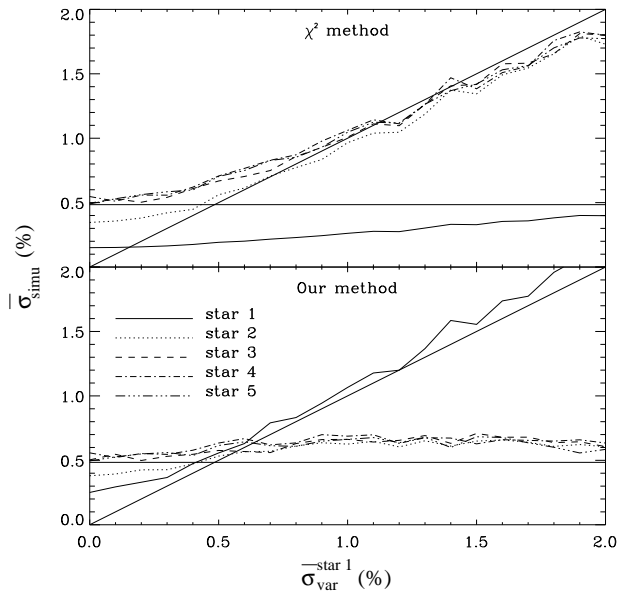}
  \caption[]{Plots of the relative standard deviations
    $\bar{\sigma}_{simu}$ of the simulated light curves of 5 stars,
    obtained with our method and the $\chi^2$ one, as a function of the
    variability amplitude $\bar{\sigma}_{var}^{star\ 1}$ of star 1. The
    other stars are only marred by statistical noise. The horizontal
    straight line in each plot gives the mean value of the real noise of
    star 2-5. The inclined straight line represents the
    $\bar{\sigma}_{simu} = \bar{\sigma}_{var}^{star\ 1}$ curve and must
    be normally followed by star 1. This is effectively the case with our
    reduction method contrary to the standard $\chi^2$
    one.\label{diffchi2}}
\end{figure}

To see the interest of our approach, let us consider a situation where at
least two stars are not variable while all the others vary independently.
Neglecting, for the moment, the statistical noise, the algorithm will
then naturally choose, for the normalization factor $N$, the common
relative flux value $N_i$ of all {\em non variable} stars, which makes
the $S_i$ function vanish. It is clearly different from the classical
minimization of the $\chi^2$ function that would give some weight to {\em
  all} stars, variable or not. However, due to the statistical noise, any
weighted algorithm will tend to favor the brightest source. This is most
apparent in the $n_s=2$ case, where the $S_i$ function reduces to the
$\chi^2$ function and the two methods become thus identical. For $n_s\geq
2$ however, they can give quite different results. We illustrate this
with a simple model: we assume that we measure 5 stars, one of which
(called star 1) is three times as luminous as each of the 4 others. We
assume that star 1 is also intrinsically variable. We simulate the light
curves of each star taking into account the statistical noise and the
intrinsic variability of star 1. Then we applied the $\chi^2$ method and
our method to the simulated data. The standard deviations of each light
curves computed by the 2 methods are plotted in Fig.  (\ref{diffchi2}) as
functions of the amplitude of the intrinsic variability of star 1.
Clearly both methods are indistinguishable when the intrinsic variability
is much lower than the mean statistical noise of stars 2-5.  However, as
soon as the variability is comparable to this value, the $\chi^2$ method
tends to underpredict the variability of star 1, because of its high
statistical weight in the normalization, and overpredict the variability
of stars 2-5. On the other hand, our method gives very good
approximations of the standard deviation of all stars.  In practice, to
use at best the advantage of our method, we choose the largest possible
number of comparison stars with approximatively the same brightness (
the relative brightness of each object can be deduced from their relative
noise $\bar{\sigma}_{ph}$ reported in Table \ref{sigtab}).  We have found
at least 3 comparison stars for all galaxies excepted NGC 4051 (only 2)
and NGC 4151 (only 1, see next).\\
For each image, the value of $N_i$ represents thus the relative flux of a
``virtual'' standard star. We finally obtain the light curve of an object
by dividing its relative flux by $N_i$.

\subsubsection{The particular case of NGC 4151}
\label{n4151}
For this object, there is only one comparison star in the CCD field with
about the same brightness as the galaxy. We obtain another comparison
object by measuring the flux of the diffuse component of NGC 4151,
excluding the central region. We have to use a large aperture and, for 
the same flux as the comparison star, the photon noise is 3
times as large due to the sky background.

\subsection{Errors measurement}
The variance of the light curve of an object depends obviously on the
method of treatment used and can be expressed, in the more general case,
as the sum of 2 terms:
\begin{equation}
\sigma_{obs}^2=\sigma^2_{ph}+\sigma^2_{supp}
\label{sigmaobs}
\end{equation}
In this expression, $\sigma_{ph}$ would be the value of $\sigma_{obs}$
obtained if the object was really non-variable and only marred by photons
statistics. On the other hand, $\sigma_{supp}$ represents a supplementary
noise which can include a variable component or any artefact of the light
curve due to the observations or the treatment. An estimation of
$\sigma_{supp}$ gives thus an estimation or an upper limit of the
variability of the object. We assess $\sigma_{supp}$ indirectly by
evaluating $\sigma_{ph}$. We simulate in this way new sets of data, where
the flux of each star $s$ in each image $i$ takes the following value:
\begin{equation}
x_{\mbox{\tiny{simu}}}^{s,i}=\left\langle x^{s,i}
  \right\rangle_{\mbox{\tiny{run}}}\left\langle\frac{x^{s,i}}{\left\langle
        x^{s,i}\right\rangle_{\mbox{\tiny{run}}}}\right
  \rangle_{\mbox{\tiny{star}}}. 
\label{simuflux}
\end{equation}
In this expression $\langle\ \rangle_{\mbox{\tiny{run}}}$ means the
average flux of a star on all the images of the run and $\langle\ 
\rangle_{\mbox{\tiny{star}}}$ means the average flux on all the stars of
an image. The second term of the right member of Eq.  (\ref{simuflux})
allows to take into account global variations of fluxes, image by image
due for example to small clouds crossing.  Finally we add a poissonian
noise to each simulated value.  Then, we treat the data with the same
algorithm described above. The standard deviation of the light curves
gives therefore an estimation of $\sigma_{ph}$ and thus, of
$\sigma_{supp}$ from Eq.(\ref{sigmaobs}). Due to the limited number of
images, there is a statistical inaccuracy on this estimation and we
improved it by repeating the simulation many times
and taking the average.\\
The value of $\sigma_{ph}$, obtained in this manner, is very close
(within a factor 2) to the true observationnal noise (photon noise and
read-out noise) and proves, by the way, the robustness of the method.

\subsection{The structure function}
A way to detect a continuous trend in our data is to used the so-called
first-order structure function (hereafter we simply refer to the
``structure function'', or ``SF''), commonly employed in time-series
analysis (Rutman \cite{Rut78}). It has been introduced in the field of
astronomy by Simonetti et al. (\cite{sim85}, see also Paltani et al.
\cite{Pal97}).  It is defined, for data of minimum temporal sampling
$\Delta t$ between two consecutive images, by:
\begin{equation}
SF_{k}(\tau=n\Delta t)=\sqrt{<{(\bar{x}_{i,k}-\bar{x}_{i+n,k})^2}>}
\end{equation}
for the star $k$ of the run. The brakets point out that we take the
average on all the images $i$ of the light curve. We can sum up the main
aspects of the structure function as follows. For a non-variable object,
the SF is constant and gives an estimation of the standard deviation of
the white noise introduced by the measurement errors on the fluxes. For
light curves with different variable components of different timescales,
the SF is more complex, increasing with $\tau$ until the maximum
variability time scale is reach.
Obviously, for small sample of images, the form of the structure
function for the largest time lags is very noisy, since the average is
done on a very small number of images.

\begin{table*}
\caption[]{Relative values of $\sigma_{obs}$, $\sigma_{ph}$ and
$\sigma_{supp}$ for each galaxy and comparison stars \label{sigtab}}
\begin{center}
\begin{tabular}{|lccc|lccc|lccc|}
\hline
 Name & $\bar{\sigma}_{obs}$ & $\bar{\sigma}_{ph}$ &
 $\bar{\sigma}_{supp}$ & Name & 
 $\bar{\sigma}_{obs}$ & $\bar{\sigma}_{ph}$ &
 $\bar{\sigma}_{supp}$ & Name & $\bar{\sigma}_{obs}$ 
 & $\bar{\sigma}_{ph}$ & $\bar{\sigma}_{supp}$\\
  & (\%) & (\%) & (\%) & & (\%) & (\%) & (\%) & & (\%) & (\%) & (\%) \\
\hline
{\bf  AKN 120}  &  0.85 &  0.62 &  0.51 &{\bf 1H 0510+031}&  1.64 &  1.82 &  0.02 &{\bf IRAS 15438+2715}&  0.89 &  0.44 &  0.74 \\   
star1 &  0.48 &  0.35 &  0.28 &  star1 &  0.53 &  0.55 &  0.22 & star1 &  0.66 &  0.36 &  0.54 \\
star2 &  0.50 &  0.28 &  0.40 &  star2 &  0.98 &  0.78 &  0.68 & star2 &  0.58 &  0.27 &  0.55 \\
star3 &  0.75 &  0.38 &  0.68 &  star3 &  1.05 &  1.25 &  0.00 & star3 &  0.75 &  0.41 &  0.67 \\
star4 &  0.86 &  0.49 &  0.72 &  star4 &  1.72 &  1.40 &  1.04 & star4 &  0.54 &  0.30 &  0.42 \\
\cline{1-4}                        
{\bf Mkn 543}  &  1.54 &  1.19 &  0.69 &  star5 &  1.52 &  1.08 &  1.17 & star5 &  0.97 &  0.49 &  0.88 \\ 
\cline{5-12} 
star1 &  2.15 &  1.19 &  1.69 &  {\bf Mkn 1044} &  0.93 &  0.68 &  0.57 & {\bf NGC 4051a}&  0.47 &  0.31 &  0.38 \\   
star2 &  1.83 &  1.52 &  0.46 &  star1 &  0.70 &  0.53 &  0.40 & star1 &  0.44 &  0.42 &  0.17 \\  
star3 &  1.60 &  1.45 &  0.33 &  star2 &  0.65 &  0.51 &  0.34 & star2 &  0.17 &  0.16 &  0.06 \\  
\cline{9-12}                    
star4 &  1.47 &  1.10 &  0.84 &  star3 &  0.84 &  0.65 &  0.50 & {\bf NGC 4051b}&  0.98 &  0.47 &  0.88 \\   
\cline{1-4}                          
{\bf Mkn 1392}  &  0.51 &  0.16 &  0.46 &  star4 &  0.75 &  0.41 &  0.69 & star1 &  0.53 &  0.60 &  0.16 \\
\cline{5-8}                         
star1 &  0.33 &  0.16 &  0.29 &  {\bf Mkn 376}  &  0.83 &  0.69 &  0.07 & star2 &  0.21 &  0.24 &  0.06 \\
\cline{9-12}
star2 &  0.36 &  0.10 &  0.39 &  star1 &  0.97 &  0.62 &  0.77 & {\bf Mkn 590}  &  0.55 &  0.43 &  0.20 \\     
star3 &  0.24 &  0.15 &  0.18 &  star2 &  0.88 &  0.85 &  0.14 & star1 &  0.81 &  0.70 &  0.43 \\  
\cline{1-4}                         
{\bf Mkn 1098}  &  0.53 &  0.24 &  0.42 &  star3 &  1.37 &  0.91 &  1.09 & star2 &  0.63 &  0.61 &  0.10 \\ 
star1 &  0.55 &  0.21 &  0.54 &  star4 &  1.53 &  1.36 &  0.63 & star3 &  0.80 &  0.61 &  0.51 \\  
star2 &  0.49 &  0.24 &  0.47 &  star5 &  1.59 &  1.29 &  1.08 & star4 &  0.80 &  0.59 &  0.55 \\
\cline{5-8}                          
star3 &  0.45 &  0.29 &  0.34 &{\bf IRAS 04448-0513}&  0.88 &  0.83 &  0.24 & star5 &  0.66 &  0.49 &  0.44 \\
\cline{9-12}                         
star4 &  0.38 &  0.19 &  0.30 &  star1 &  0.76 &  0.59 &  0.52 &{\bf PG 0844+349}&  1.00 &  0.88 &  0.48 \\    
\cline{1-4}                          
{\bf Mkn 335}  &  1.20 &  0.96 &  0.53 &  star2 &  0.67 &  0.62 &  0.21 & star1 &  0.63 &  0.57 &  0.30 \\
star1 &  0.67 &  0.65 &  0.03 &  star3 &  0.76 &  0.74 &  0.15 & star2 &  1.75 &  1.58 &  0.69 \\  
star2 &  0.61 &  0.53 &  0.16 &  star4 &  0.80 &  0.56 &  0.62 & star3 &  0.84 &  0.81 &  0.27 \\  
star3 &  0.63 &  0.46 &  0.42 &  star5 &  0.66 &  0.50 &  0.46 & star4 &  1.75 &  1.66 &  0.72 \\
\cline{1-8}                          
{\bf Mkn 478}  &  0.53 &  0.26 &  0.44 &  {\bf NGC 1019} &  0.61 &  0.45 &  0.34 & star5 &  0.65 &  0.48 &  0.46 \\
star1 &  0.23 &  0.11 &  0.21 &  star1 &  0.62 &  0.49 &  0.28 & star6 &  0.56 &  0.43 &  0.37 \\
\cline{9-12}                         
star2 &  0.56 &  0.27 &  0.52 &  star2 &  0.85 &  0.53 &  0.64 & {\bf Mkn 359} &  0.95 &  0.30 &  0.97 \\    
star3 &  0.40 &  0.20 &  0.39 &  star3 &  0.70 &  0.54 &  0.47 & star1 &  0.25 &  0.24 &  0.14 \\  
\cline{1-4}                          
{\bf MCG+08-11-11a}&  0.54 &  0.39 &  0.35 &  star4 &  0.46 &  0.40 &  0.02 & star2 &  0.64 &  0.52 &  0.55 \\
star1 &  0.44 &  0.44 &  0.23 &  star5 &  0.90 &  0.69 &  0.65 & star3 &  1.65 &  0.52 &  1.85 \\  
star2 &  0.52 &  0.40 &  0.36 &  star6 &  1.01 &  0.60 &  0.82 & star4 &  0.53 &  0.40 &  0.40 \\
\cline{5-8}                          
star3 &  0.46 &  0.40 &  0.23 &  {\bf Mkn 684}  &  0.93 &  0.25 &  0.89 & star5 &  0.95 &  0.39 &  0.97 \\   
star4 &  0.53 &  0.37 &  0.38 &  star1 &  0.57 &  0.13 &  0.65 & star6 &  0.75 &  0.43 &  0.69 \\   
\cline{1-4} 
{\bf MCG+08-11-11b}&  0.47 &  0.41 &  0.17 &  star2 &  0.45 &  0.16 &  0.17 & star7 &  0.53 &  0.20 &  0.53 \\
\cline{9-12}                            
star1 &  0.32 &  0.46 &  0.00 &  star3 &  0.86 &  0.25 &  0.94 & {\bf NGC 4151} &  0.63 &  0.18 &  0.56 \\
\cline{5-8}
star2 &  0.60 &  0.41 &  0.45 &  {\bf Mkn 1383} &  0.51 &  0.15 &  0.45 & star1 &  1.05 &  0.52 &  0.99 \\   
star3 &  0.44 &  0.42 &  0.09 &  star1 &  0.48 &  0.22 &  0.44 & star2 &  0.09 &  0.03 &  0.30 \\  
star4 &  0.50 &  0.40 &  0.29 &  star2 &  0.41 &  0.27 &  0.31 & & & & \\                            
\cline{1-4}                                                   
{\bf Mkn 372}  &  0.78 &  0.60 &  0.48 &  star3 &  0.29 &  0.13 &  0.29 & & & & \\ 
\cline{5-8}
star1 &  0.50 &  0.44 &  0.29 &  {\bf Mkn 9}  &  0.88 &  0.80 &  0.16 & & & & \\                            
star2 &  0.50 &  0.46 &  0.19 &  star1 &  1.24 &  0.91 &  0.99 & & & & \\                            
star3 &  0.56 &  0.41 &  0.38 &  star2 &  0.97 &  0.72 &  0.61 & & & & \\                            
star4 &  0.77 &  0.73 &  0.32 &  star3 &  0.99 &  0.50 &  0.94 & & & & \\                            
star5 &  0.72 &  0.65 &  0.31 &  star4 &  0.72 &  0.67 &  0.17 & & & & \\                              
      &       &       &       &  star5 &  1.19 &  0.71 &  0.93 & & & & \\
\hline
\end{tabular}
\end{center}
\end{table*}

\section{Variability results}
\label{varres}
\subsection{Optical observations}
\begin{figure*}[tbp]
\begin{tabular*}{\textwidth}{@{\excs}ll@{\extracolsep{0pt}}}
\includegraphics[width=\columnwidth]{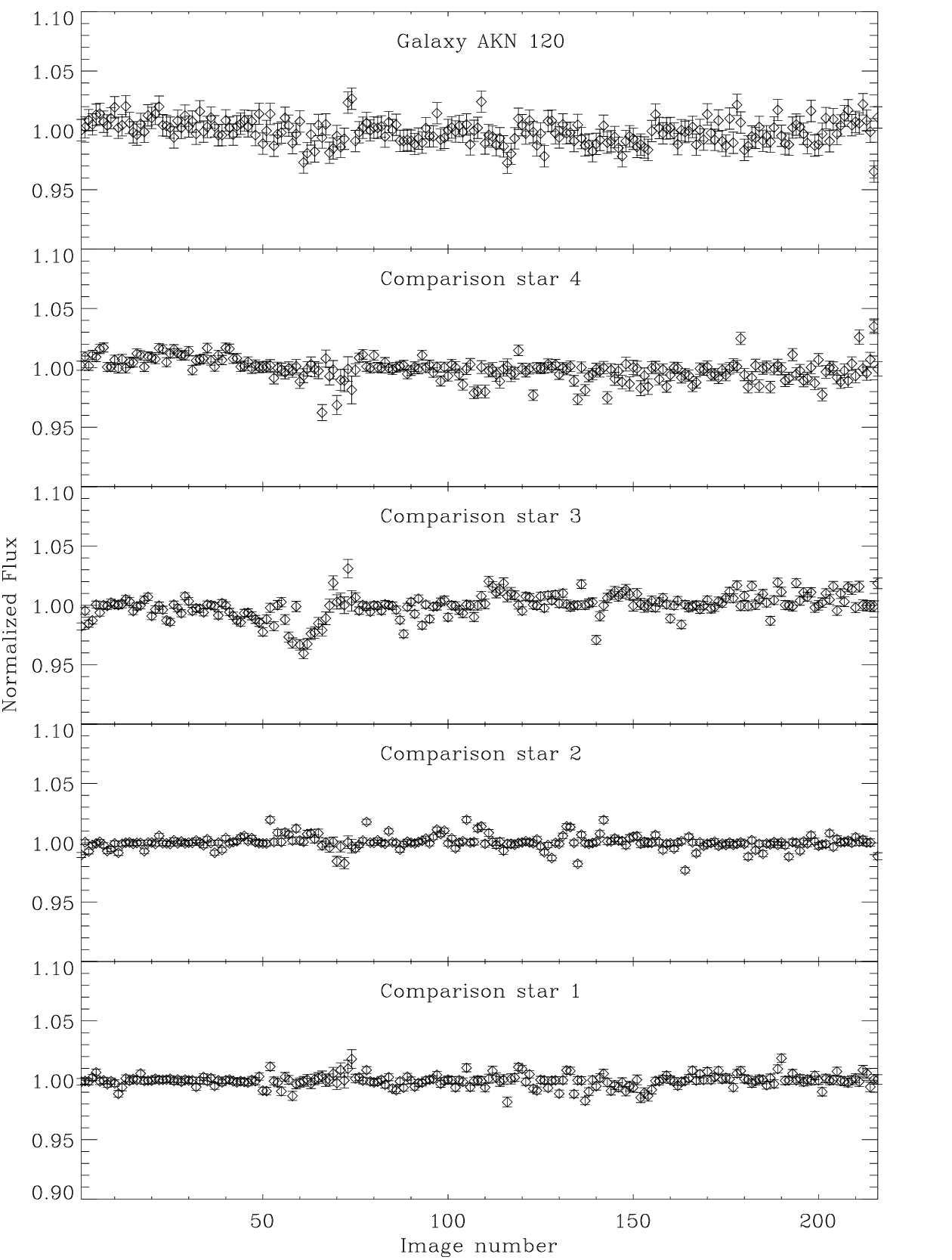} & 
\includegraphics[width=\columnwidth]{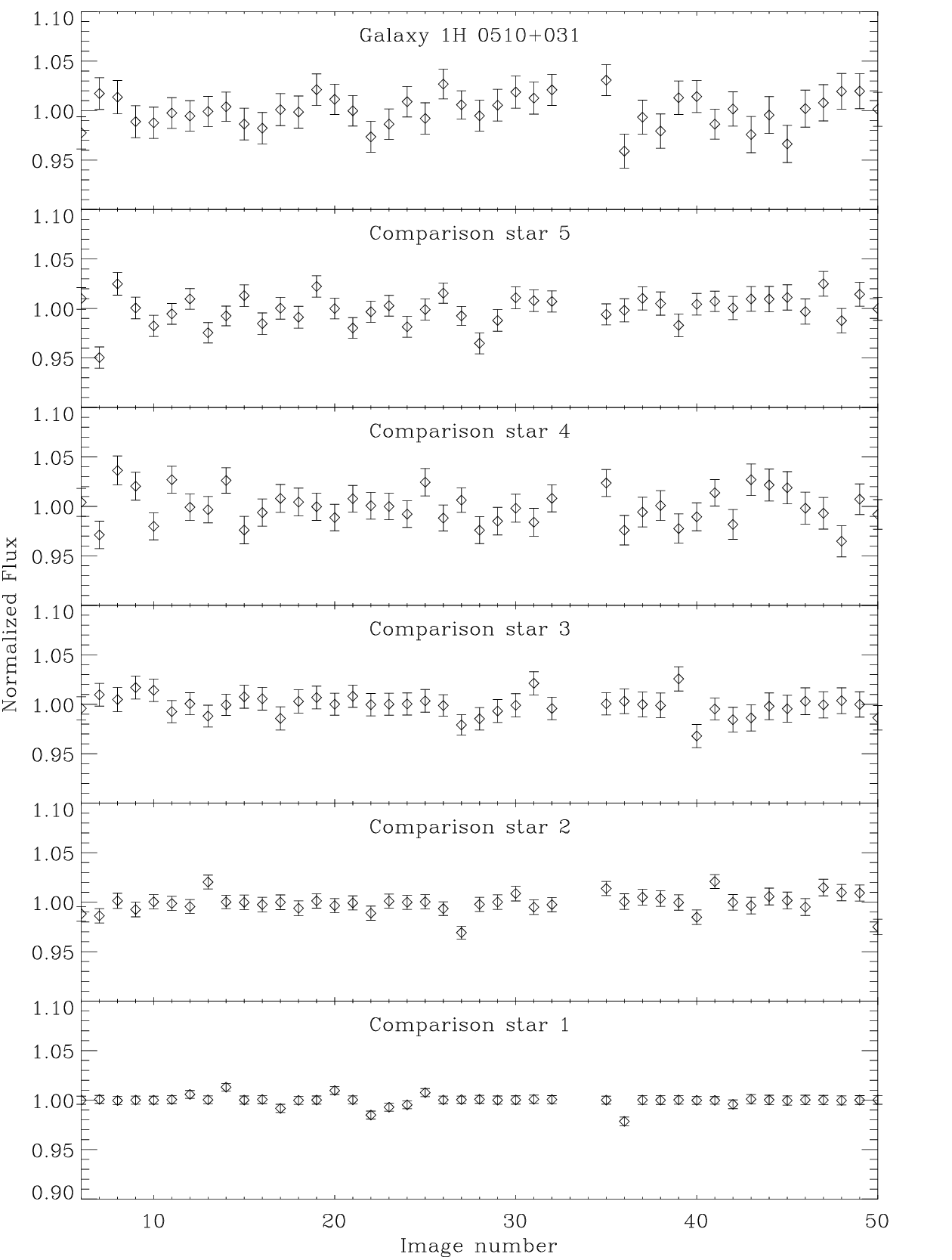} \\
\includegraphics[width=\columnwidth]{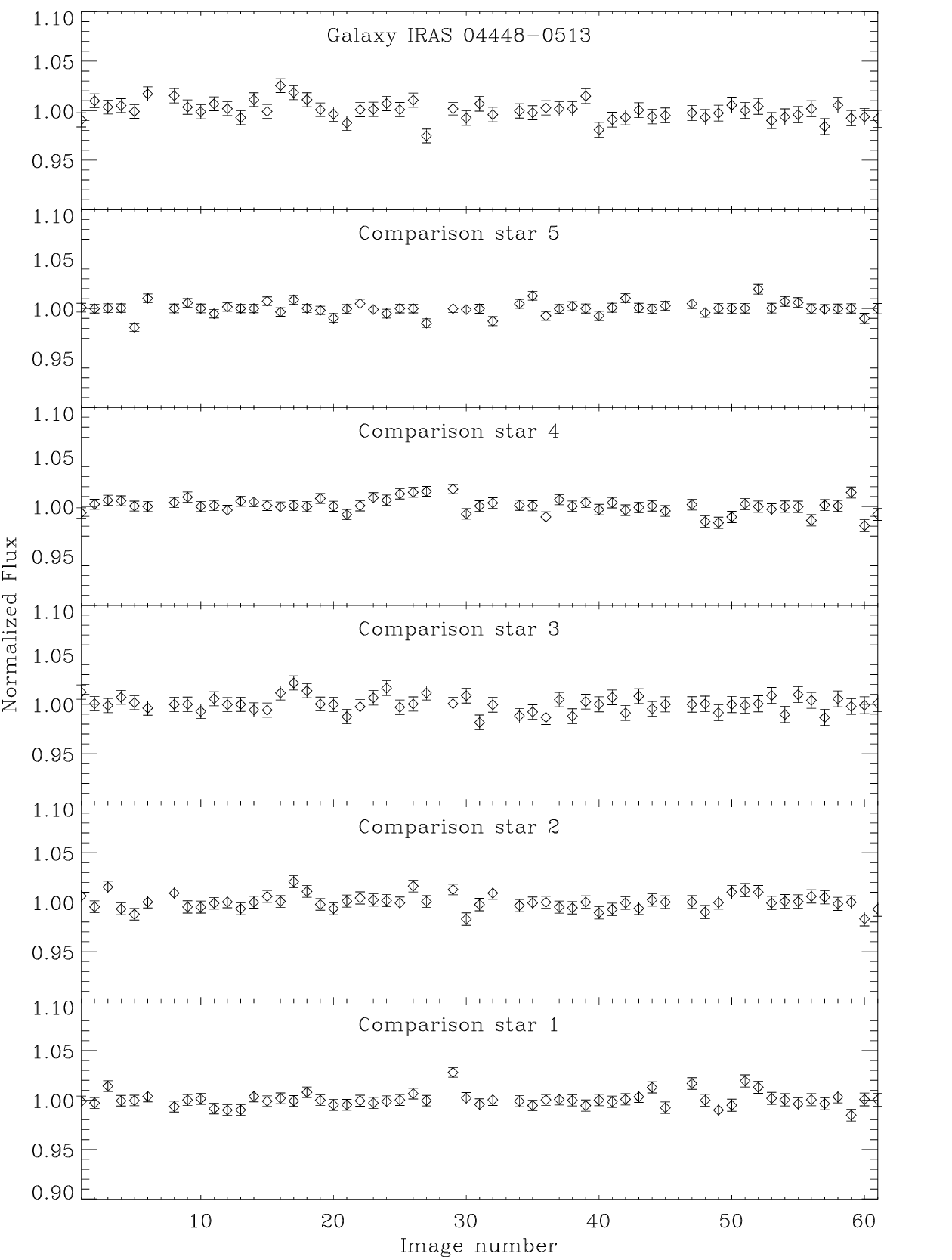} &
\includegraphics[width=\columnwidth]{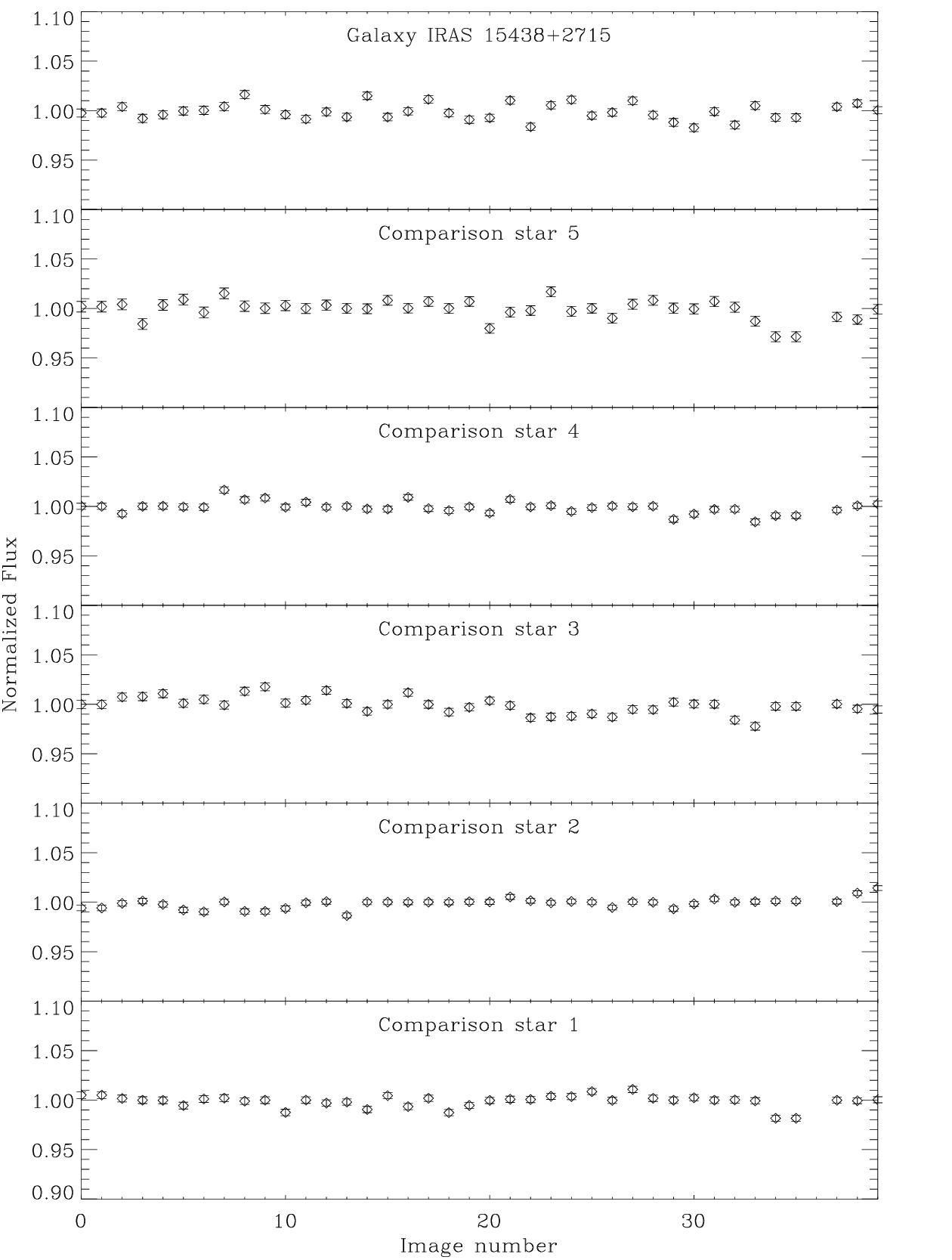} \\
\end{tabular*}
\caption[]{Light curves of the different galaxies and associated
  comparison stars of the sample.\label{lightcurves}  }
\end{figure*}

The relative values of $\sigma_{obs}$, $\sigma_{ph}$ and
$\sigma_{supp}$ are reported in table (\ref{sigtab}) for each galaxy and
comparison stars. Values of $\sigma_{supp}$ are smaller than $1\%$ in
\begin{figure}[tbp]
\includegraphics[angle=0,width=\columnwidth,origin=br]{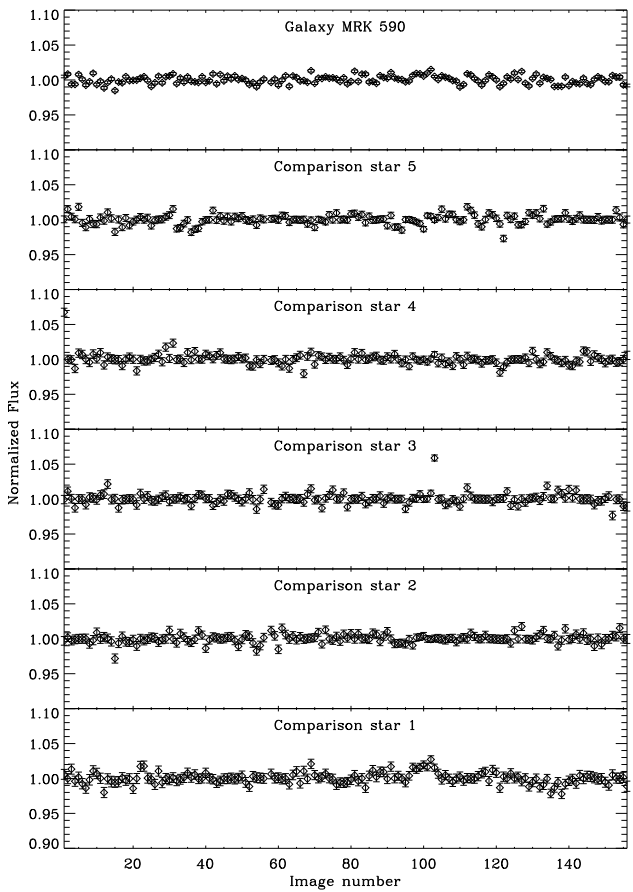}
  \caption[]{The typical light curves of a non-variable (during our
    observations) Seyfert galaxy (here Mkn 590) and 5 comparisons stars
    used for differential photometry.}\label{m590lightcurve}
\end{figure}
\begin{figure}[tbp]
\includegraphics[angle=0,width=\columnwidth,origin=br]{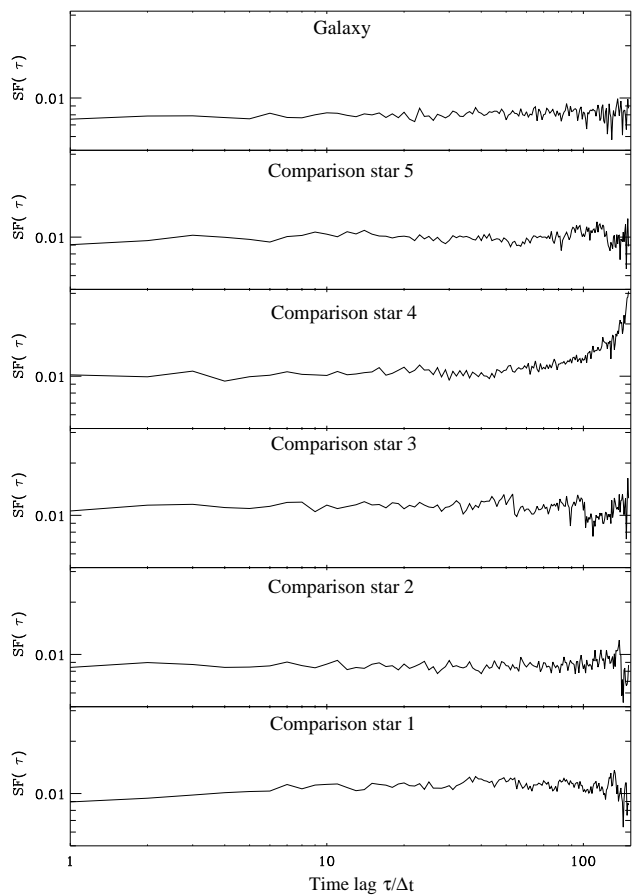}
  \caption[]{The corresponding structure functions of Mkn 590 and its
    associated comparisons stars with a temporal sampling $\Delta t\simeq
    90$ s. No variability time lag appears.}\label{m590structfunc}
\end{figure}
all cases and often smaller than $0.5\%$ which underlines the precision
of the method employed. We consider a galaxy to be variable if firstly
$\sigma_{obs}\geq 2\sigma_{ph}$ (i.e. $\sigma_{supp}\geq
\sqrt{3}\sigma_{ph}\simeq 2\sigma_{ph}$) and secondly at least one
comparison star is stable. From table (\ref{sigtab}), it appears that we
have no clear variability detection for any of the objects of the sample,
with some limited cases for Mkn 478, Mkn 684, Mkn 1392 and NGC 4151,
studied further. Figure \ref{m590lightcurve} shows the typical light
curves obtained by our algorithm for a non-variable (according to the
previous criteria) galaxy, Mkn 590, and 5 comparison stars. The
corresponding structure functions are also plotted in Fig.
\ref{m590structfunc}. They are all flat (the form of the structure
function for $\tau \ge 100$ is smarred by large statistical errors not
plotted in the graph) meaning that no continuous trend are present in the
data during the period of observations. The light curves of each galaxy
and the associated comparison stars are plotted in Fig.
\ref{lightcurves} at the end of this paper.

\subsection{Individual objects}
We only presents results for the most interesting objects either because
they have a limit variability detection or because they have been
previously studied by other authors. We develop succinctly some tests
used to confirm or not any variability detection.

\subsubsection{Mkn 684 and Mkn 1383}
These galaxies fulfill the two criteria of variability since
$\sigma_{supp}\geq 3 \sigma_{ph}$ and their comparison star 2 is non
variable.  Yet only one comparison star, in each case, has a flat
structure function and the other ones increase with time lag. We suspect
that a selection effect may occur in our algorithm (see Section
\ref{treatsection}). Thus, we start again the treatment, including the
galaxy in the set of comparison stars. All the new structure functions
appear finally stable for all time lags, quashing any variability
detection.

\subsubsection{Mkn 1392}
\begin{figure}[tbp]
\includegraphics[angle=0,width=\columnwidth,origin=br]{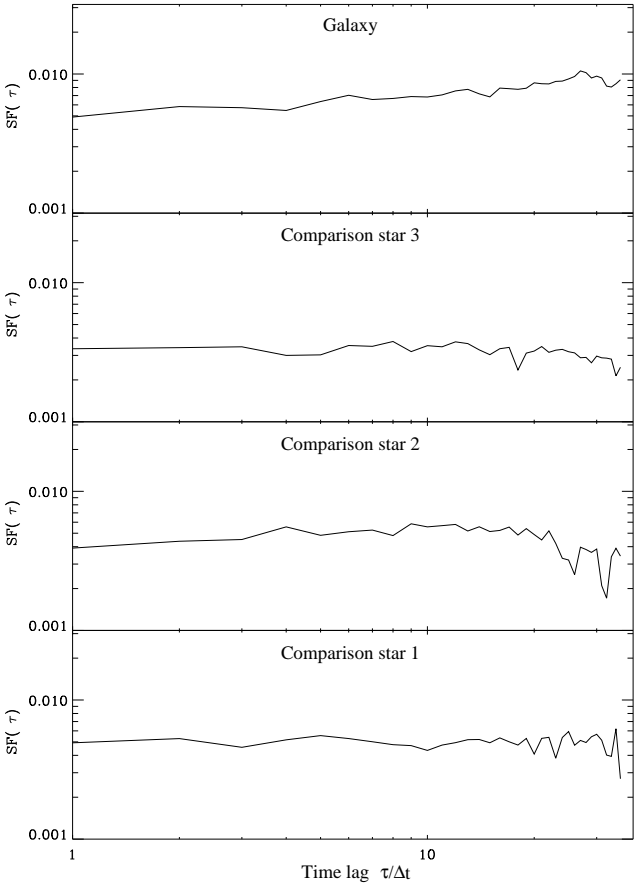}
  \caption[]{The corresponding structure functions of Mkn 1392 and its
    associated comparisons stars with a temporal sampling $\Delta t\simeq
    300$ s. The structure function of the galaxy increases slightly
    during the run which may indicate a variability on time scale larger
    than the length of the observation.}
\label{m139structfunc}
\end{figure}
This galaxy fullfills equally the criteria of variability and its
structure function increases slightly during the run whereas the
comparison star ones are stable (see Fig. \ref{m139structfunc}). This
trend remains even if we include the galaxy in the comparison star. It
seems, thus, that Mkn 1392 may be variable but on a timescale larger than
the length of the observations ($\simeq$ 4 hours).

\subsubsection{NGC 4151}
As previously said (see Section \ref{n4151}), important selection effect
may exist in the treatment of this galaxy, since there is only one brigth
star in the CCD field. To minimize these effects, we repeat the treatment
but including the galaxy in the set of comparison stars. The structure
function of the galaxy and its diffuse component become stable whereas
the star one slightly increases during the run. We have thus to be very
carefull when using differential photometry with this galaxy, since the
nearest bright star seems to be variable on timescale of hours.

\subsection{Infrared observations}
\begin{table}
\caption[]{Same as table \ref{sigtab} but in J band for 3 galaxies of the
  sample \label{sigtabIR}}
\begin{center}
\begin{tabular}{lccc}
\hline
 Name & $\bar{\sigma}_{obs}$ & $\bar{\sigma}_{ph}$ &
 $\bar{\sigma}_{supp}$ \\ 
 & (\%) & (\%) & (\%) \\
\hline
{\bf Mkn 1392} &  3.55 &  1.59 &  2.30 \\
star1 &  2.68 &  0.95 &  3.49 \\
star2 &  3.86 &  2.32 &  2.40 \\
star3 &  3.02 &  1.92 &  0.90 \\
star4 &  6.63 &  3.30 &  6.05 \\
\hline
{\bf Mkn 478} &  5.12 &  1.59 &  5.12 \\
star1 &  2.26 &  0.62 &  2.07 \\
star2 &  6.69 &  1.95 &  6.43 \\
\hline
{\bf Mkn 1098} &  2.71 &  1.40 &  2.30 \\
star1 &  4.31 &  1.40 &  3.80 \\
star2 &  0.70 &  1.00 &  3.50 \\
star3 &  3.02 &  2.80 &  0.60 \\
\hline
\end{tabular}
\end{center}
\end{table}
Three galaxies of the sample, Mkn 478, Mkn 1392 and Mkn 1098, have been
observed simutaneously in I and J bands. We treat the J band data with
the same algorithm described in Section \ref{treatsection}. At this
wavelength we are clearly limited by the CCD and sky background noises.
We can not obtain precision smaller than 2\% and it is in the range 2-5
\% in most cases. The results are reported in table \ref{sigtabIR}. Only
Mkn 478 fulfills the first variability criterium, since all comparison
stars seem variable. But, once again, only two comparisons stars were
used and a selection effect occurs. No more variability is detected when
the galaxy is included in the set of comparison stars.

\section{Discussion}
\label{physicalconstraint}
The simultaneous observations of NGC 4051 in the IR-optical and X-ray
wavebands by Done et al (\cite{Don90}) have given very strong constraints
on the spatial distribution of the emitting regions.  Effectively, in
this object, the limits on the amount of rapid variability in the
optical/IR were below 1 and 5 per cent while the X-ray flux continually
flickered by up to a factor 2. It clearly rules out models in which the
IR/optical and X-ray continuum emission are produced in the same region.
Nonetheless, the IR/optical continuum could be the sum of two different
components.  The first one could originate in the outflows observed in
most Seyfert galaxies (Wilson \cite{Wil93}, Colbert et al. \cite{Col96}),
through synchrotron process on large scale magnetic field. Due to the
large sizes of the flows, we expect no rapid variabilities from this
emission. On the contrary, a second component, whose flux is noted
$F_{syn}$, could be associated with the synchrotron emission of the
non-thermal distribution of relativistic electrons producing X-rays, and
thus concentrated in a much smaller region. Since rapid X-ray variability
is a common features in such objects (Mc Hardy et al.  \cite{McH85},
Mushotzky et al. \cite{Mus93}, Grandi et al.  \cite{Gra92}) and is likely
associated with instabilities in the source of particles, we expect
flickering from this second component too. We assume that its variability
amplitude is of the order of the flux, that is $\sigma(F_{syn})\simeq
F_{syn}$ which seems reasonable since it is the case in the X-ray range
(Mushotzky et al. \cite{Mus93}, Ulrich et al. \cite{ulr97}). The
treatment allows thus to estimate an upper limit of this variable
component by measuring $\sigma_{supp}$ and therefore to constrain the
intrinsic properties of the local environment of the emission region. Our
assumptions are presented in the following.

\subsection{Basic hypotheses}
\begin{table*}[tbp]
\begin{center}
\caption[]{Characteristics of 7 galaxies of the sample. The flux density
  are given in $10^{-11}erg.s^{-1}.cm^{-2}$, lengths in centimeter and
  magnetic fields in gauss units. Data are taken from Walter \& Fink
  \cite{Wal92}.  The maximum of $B_{\mbox{\tiny{sup}}}$ gives an
  absolute upper limit on the magnetic field in the AGN in order not to
  detect variability.\label{tab3}}
\begin{tabular}{lccccccc}
\noalign{\smallskip}
\hline
\noalign{\smallskip}
 Name & $F^{\mbox{\tiny{2 keV}}}_{\mbox{\tiny{X}}}$ &
 $F^{\mbox{\tiny{1375 \AA}}}_{\mbox{\tiny{UV}}}$ & $F_{\mbox{\tiny{syn}}}$ & s & $R_s$ & $R_{var}$ & $max(B_{\mbox{\tiny{sup}}})$ \\  
\noalign{\smallskip}
\hline
\noalign{\smallskip}
Mkn 359        & $0.12\pm 0.01$ & $2.76\pm 0.54$ & 0.03 & 0.9  & 8$10^{10}$ & - & 2.5$10^5$\\
Mkn 590        & $1.51\pm 0.22$ & $8.36\pm 1.08$ & 0.01 & 0.9  & 6$10^{11}$ & - & 1.6$10^4$\\
ARK 120        & $1.56\pm 0.06$ & $22.3\pm 3.3 $ & 0.03 & 1.1  & 2$10^{12}$ & - & 1.9$10^4$\\
MCG+08-11-11   & $2.55\pm 0.11$ & $1.98\pm 2.46$ & 0.01 & 0.7  & 8$10^{10}$ & - & 8000       \\
NGC 4051       & $0.54\pm 0.03$ & $1.66\pm 0.44$ & 0.17 & 0.65 & 6$10^8$& 9$10^{12}$ & 4.7$10^6$\\
Mkn 1383       & $0.42\pm 0.03$ & $11.6\pm 3.9 $ & 0.12 & 0.9  & 8$10^{12}$ & - & 2.4$10^4$\\
Mkn 478        & $0.41\pm 0.04$ & $7.47\pm 2.04$ & 0.008& 0.9  & 4$10^{12}$ & - & 9.0$10^3$\\
\noalign{\smallskip}
\hline
\end{tabular}
\end{center}
\end{table*}
We suppose the non-thermal plasma region to be spherical, with radius
$R$. As explained above, the particles emit synchrotron radiation in a
magnetic field of strength $B$. We also assume the electrons density
distribution follows a power law with spectral index $s$, i.e.
$\displaystyle n(\gamma)=n_0 \gamma^{-s}$, with
$\gamma_{min}\le\gamma\le\gamma_{max} (\gg \gamma_{min})$. If we assume
the magnetic field to be uniform throughout the emitting region and with
a random direction in the line of sight, the spectral density of the
synchrotron flux received by an observer at a distance $D$ away, can be
approximated by (Blumenthal \& Gould \cite{Blu70}):
\begin{equation}
\label{eqsyn}
F_{\nu}^{syn}= \left\{ \begin{array}{ll}
     E(s)B^{\frac{s+1}{2}} \frac{n_0R^3}{D^2}\nu^{-\frac{s-1}{2}} & \mbox{if
       $\nu_t < \nu < \nu_c$}\\ 
     0 & \mbox{if $\nu_c < \nu$}
               \end{array}
               \right.
\end{equation}
In this equation, $E(s)$ is a function of $s$ solely, $\nu_c$ the cut-off
frequency of the radiation which depends on the maximum Lorentz factor of
the electrons (Blumenthal \& Gould \cite{Blu70}, Rybicki \& Lightman
  \cite{Ryb89}):
\begin{equation}
\nu_c\simeq\frac{3q}{4\pi mc}\gamma_{max}^2B.
\end{equation}
and $\nu_t$ the synchrotron self-absorption frequency separating the
optically thin and optically thick regimes of synchrotron emission
(see Pacholczyk \cite{Pac70}). \\
On the other hand, the same electron population produces X-ray radiation
by Inverse Compton (IC) process on UV photons, generally supposed to be
produced by an accretion disk . We assume that the UV source
is roughly at a distance $Z$ from the non-thermal plasma. Finally we
suppose that the UV photons density can be approximate by a delta
function, and thus, at the location of the hot source, this density can
be expressed as follows:
\begin{equation}
n_{\mbox{\tiny{UV}}}(\nu)=\frac{F_{\mbox{\tiny{UV}}}D^2}
 {h\nu_{\mbox{\tiny{UV}}} Z^2c}\delta(\nu-\nu_{\mbox{\tiny{UV}}}) 
\end{equation}
where $F_{\mbox{\tiny{UV}}}$ is the observed UV flux. We can then deduced
the X-ray flux received by an observer at a distance $D$ away (Blumenthal
\& Gould \cite{Blu70}, Rybicki \& Lightman \cite{Ryb89}):
\begin{equation}
\label{eqIC}
F_{\nu}^{\mbox{\tiny{X}}}=F(s)F_{\mbox{\tiny{UV}}}\frac{n_0 R^3}
 {Z^2}\nu_{\mbox{\tiny{UV}}}^{\frac{s-3}{2}}\nu^{-\frac{s-1}{2}}.
\end{equation}
where $F(s)$ is solely a function of $s$. This expression is
representative of the common spectrum of Seyfert galaxies between 2-10
keV which is well fitted by a power with mean spectral index $s\simeq2.8$
(Mushotzky et al. \cite{Mus93}).

\begin{figure*}[tbp]

\begin{tabular*}{\textwidth}{@{\excs}ll@{\extracolsep{0pt}}}
\includegraphics[width=0.5\textwidth]{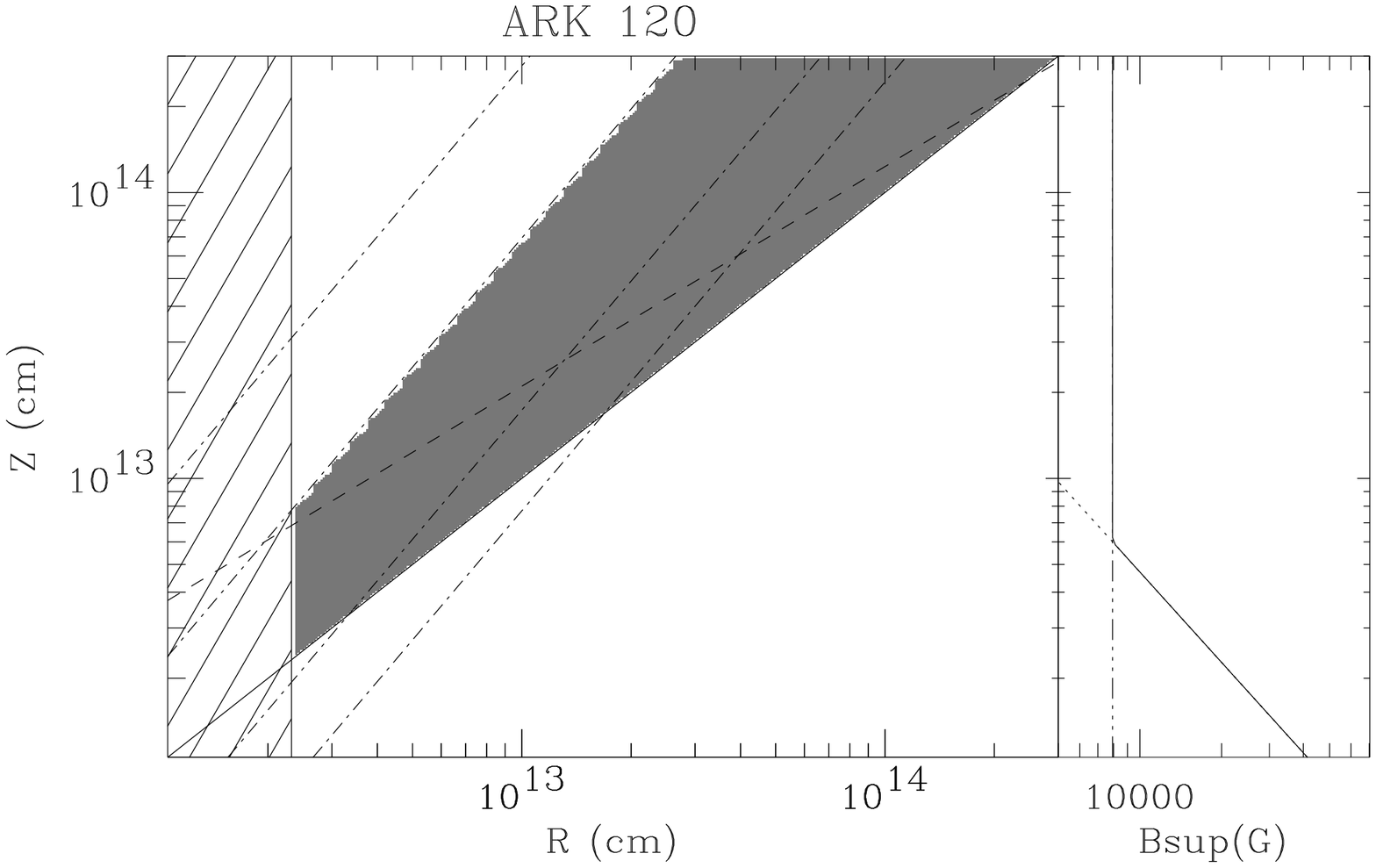} & 
\includegraphics[width=0.5\textwidth]{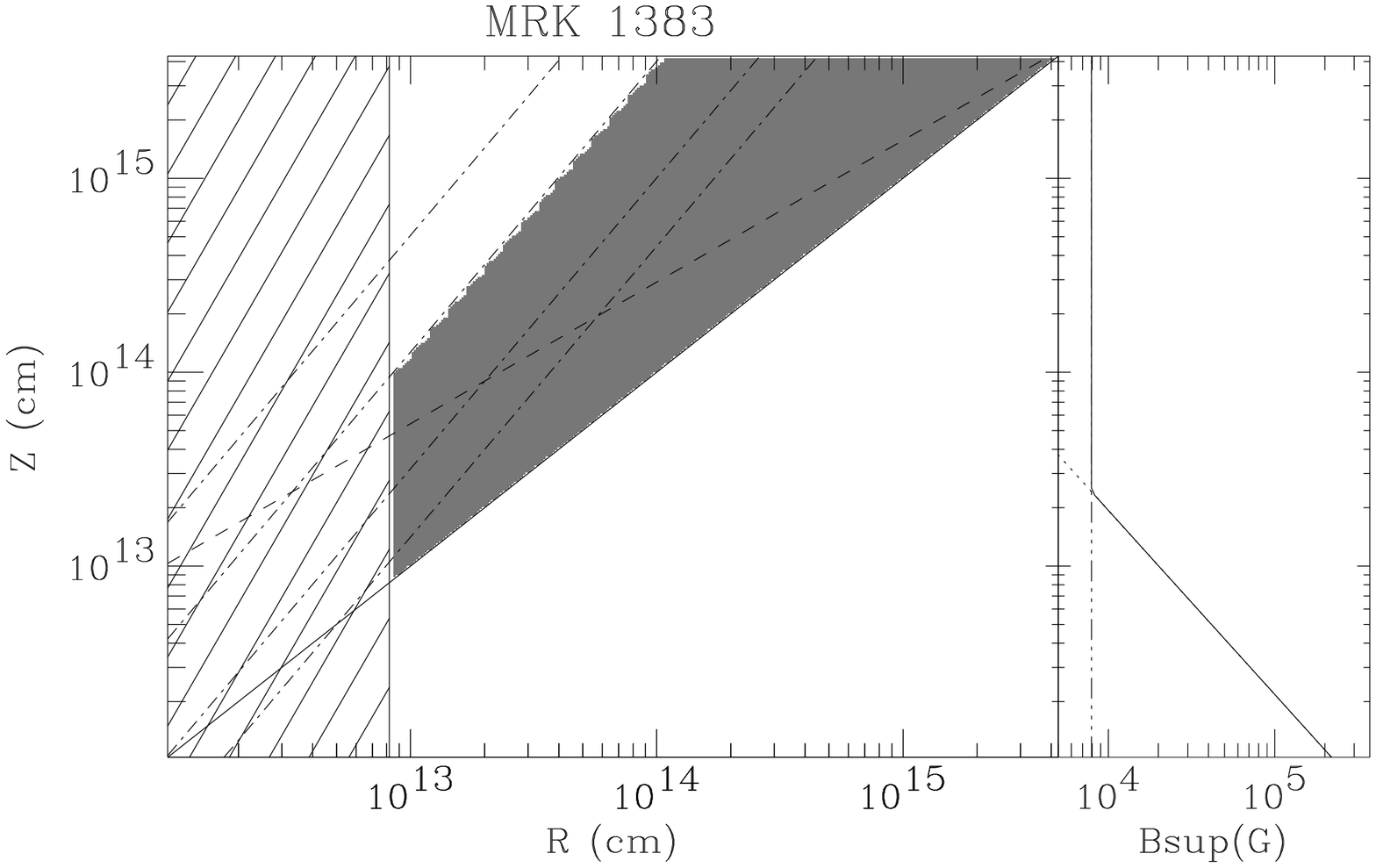} \\
\includegraphics[width=0.5\textwidth]{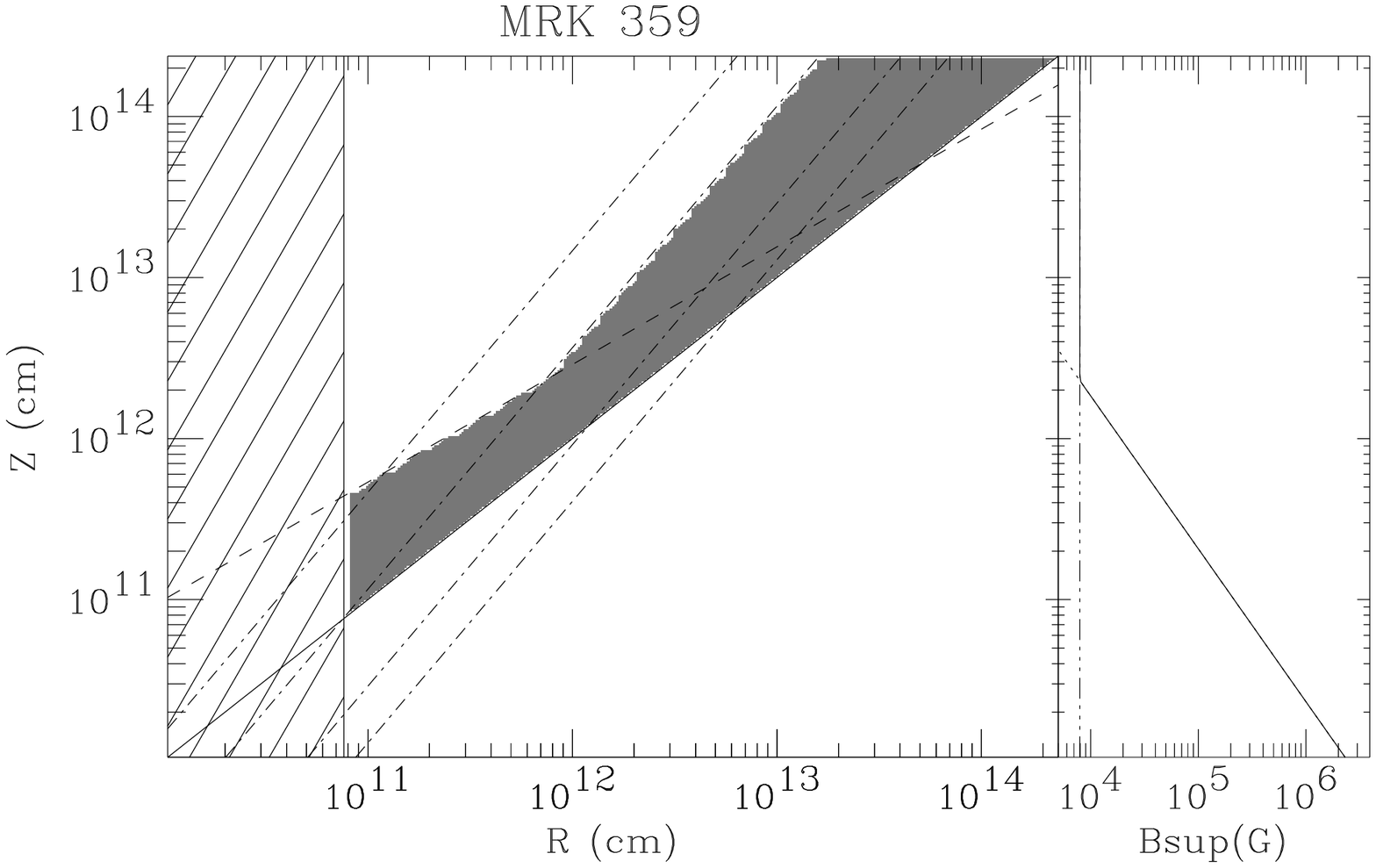} &
\includegraphics[width=0.5\textwidth]{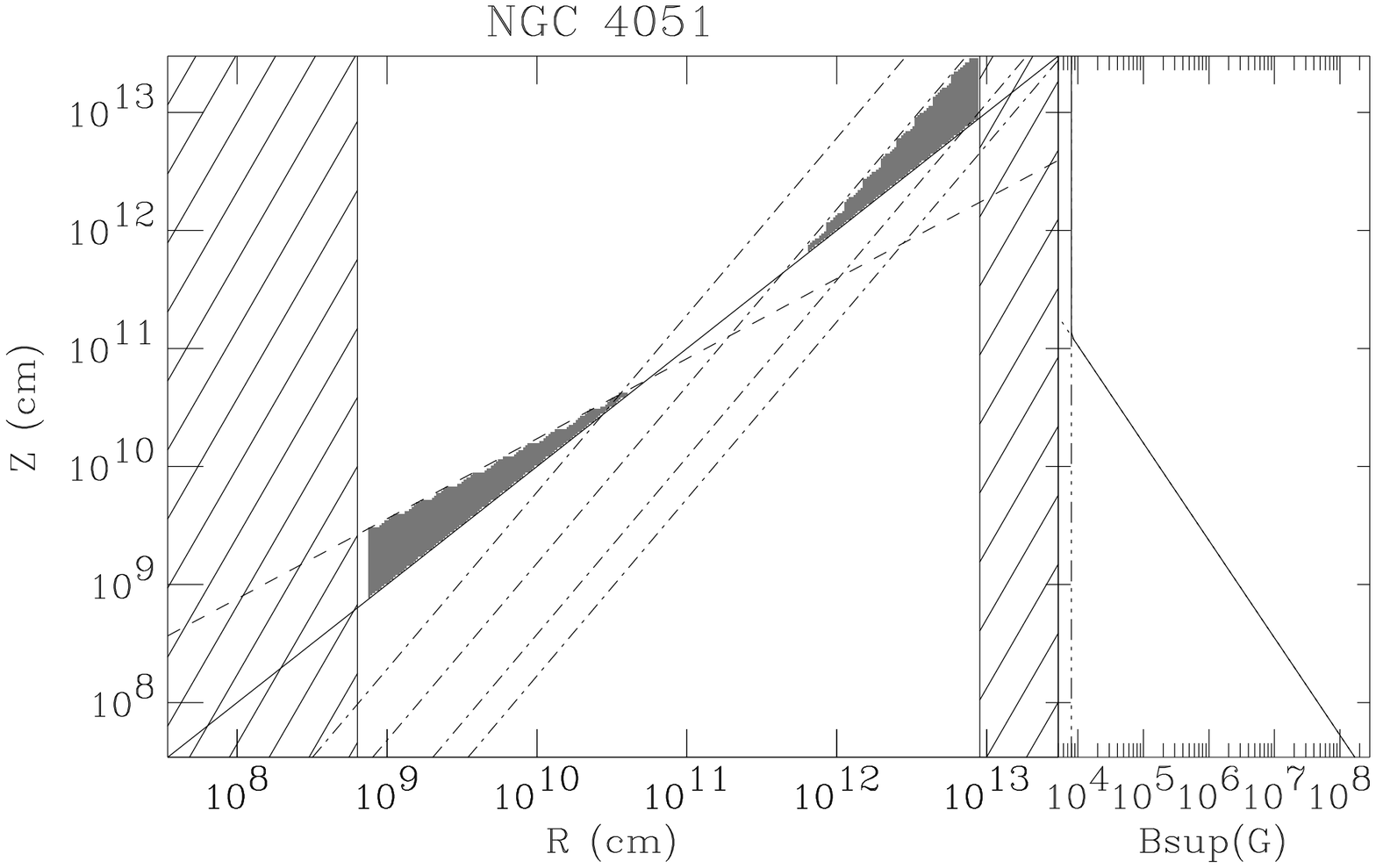} \\
\includegraphics[width=0.5\textwidth]{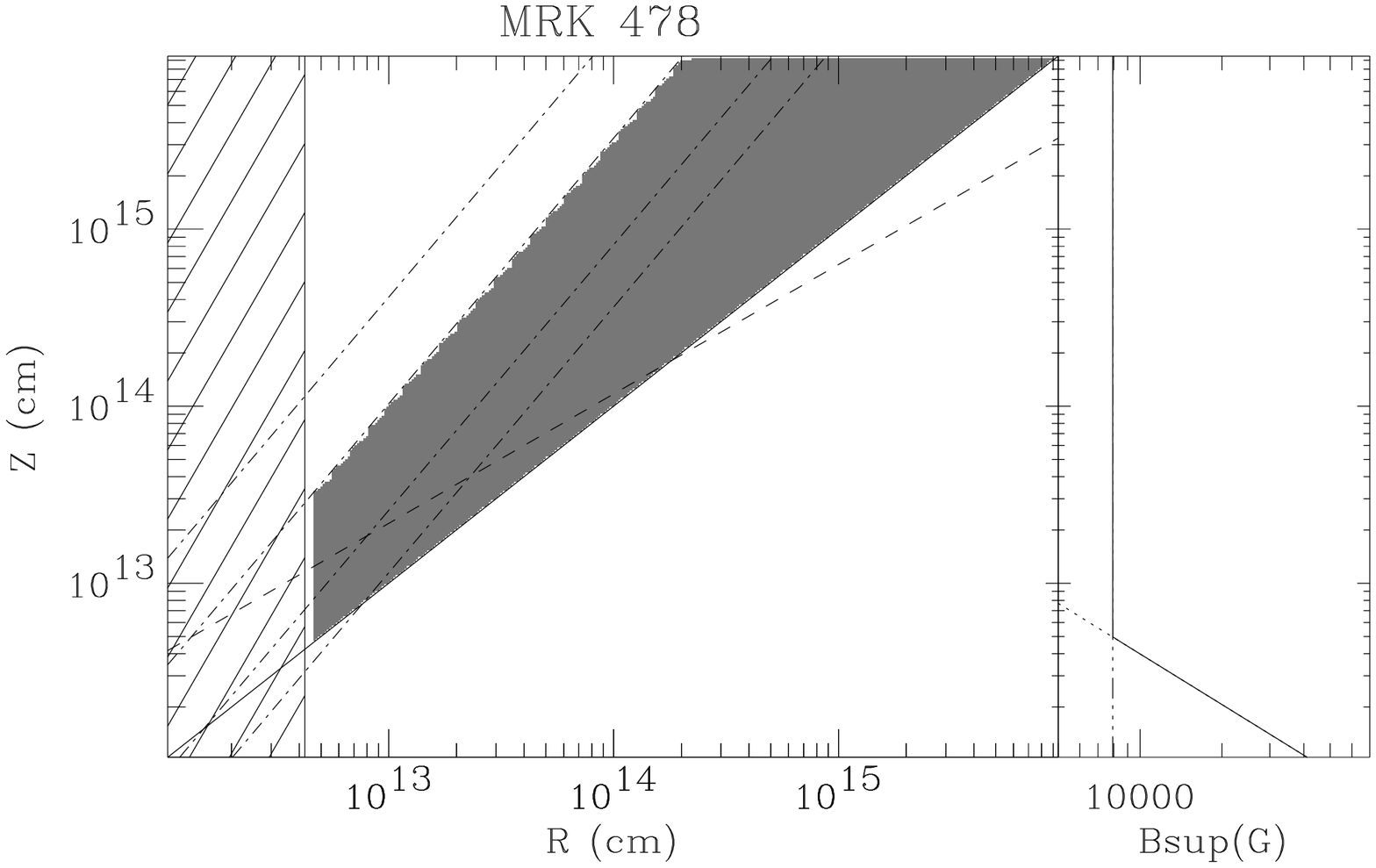} &
\includegraphics[width=0.5\textwidth]{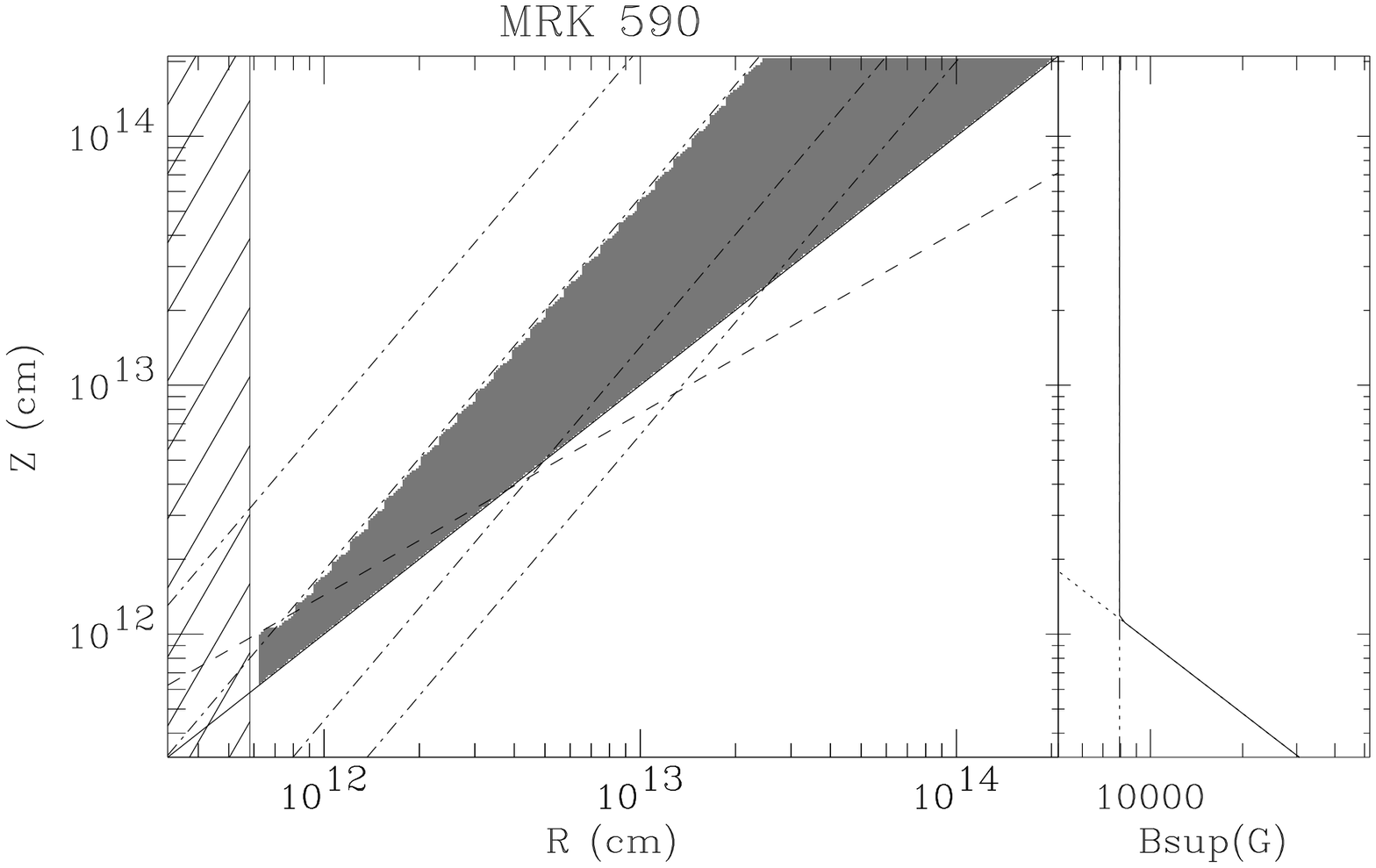} \\
\includegraphics[width=0.5\textwidth]{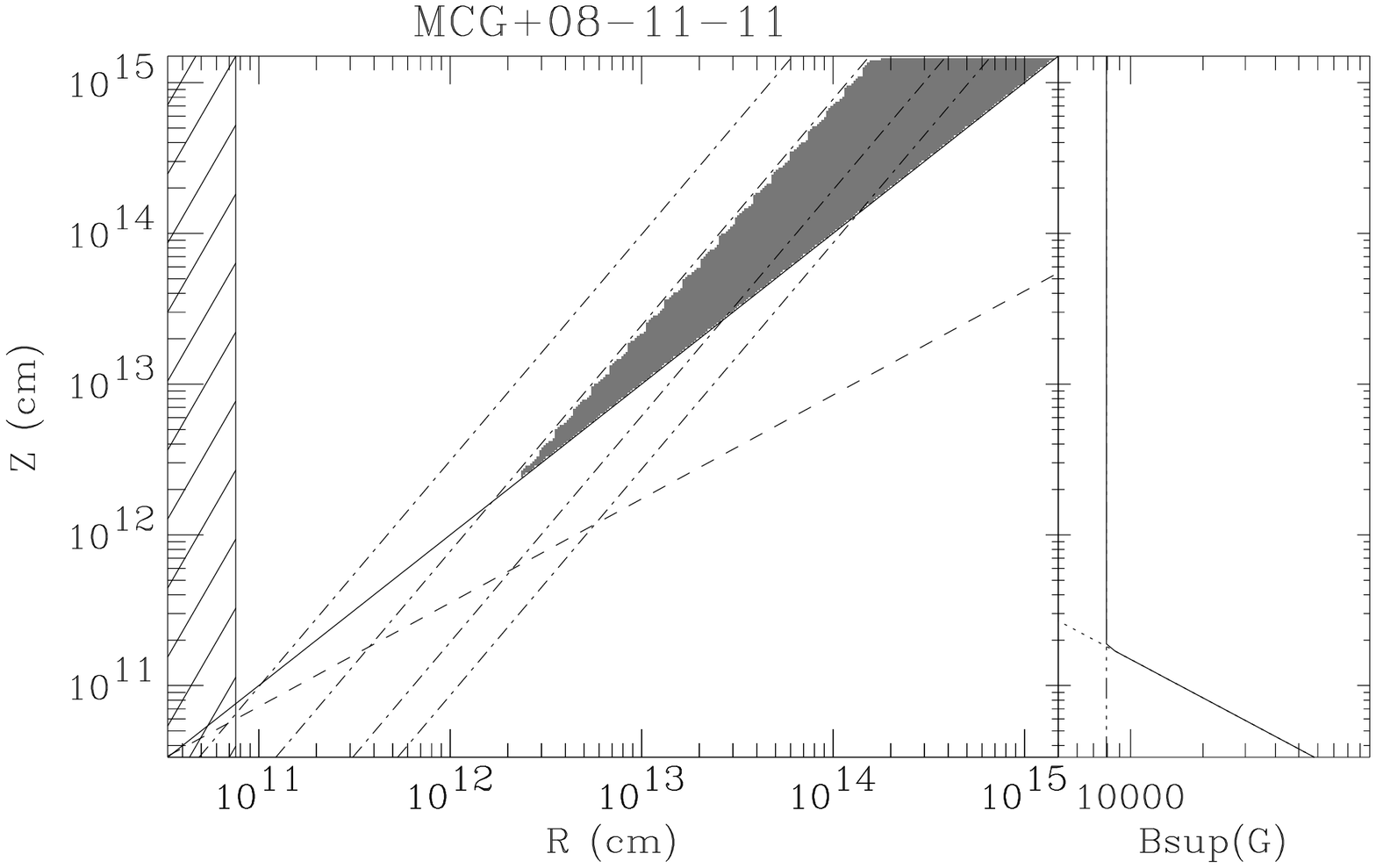} & \\
\end{tabular*}
\caption[]{\label{ZvsR} The left part of each plot gives limits on R
  and Z for 7 galaxies of our sample whose parameters are reported in
  Table \ref{tab3}. The dash lines represent the equipartition
  $B_{eq}=B_{\mbox{\tiny{flux}}}$ (Eq.  (\ref{Z1})) whereas the set of
  dot-dash lines represents the equipartition
  $B_{eq}=B_{\mbox{\tiny{cut-off}}}(\gamma_{max})$ for different values
  of $\gamma_{max}$ (Eq. (\ref{Z2})). From left to right, $\gamma_{max}$=
  50, 100, 200 and 300, corresponding to $B_{eq}\simeq$ 32000~G, 8000~G,
  2000~G and 1000~G.  Finally, the solid line refers to the $Z=R$ (Eq.
  (\ref{ZsupR})). For each galaxy, the allowed region is constrained by
  Eqs. (\ref{Rmin}), (\ref{Rmax}), (\ref{ZsupR}), (\ref{Z1}) and
  (\ref{Z2}).  It is colored in grey in each plot for $\gamma_{max}=100$.
  Other $\gamma_{max}$ values would correspond to another dot-dash curve
  (called type II in the text).  The hashed
  regions are forbidden by Eqs. (\ref{Rmin}) and (\ref{Rmax}).\\
  On the right part of each graphic, we have plotted
  $B_{\mbox{\tiny{sup}}}(Z)$ in solid line. The dot line and three
  dots-dash line correspond respectively to $B_{\mbox{\tiny{flux}}}(Z)$
  and $B_{\mbox{\tiny{cut-off}}}(Z)$. Thus
  $B_{\mbox{\tiny{sup}}}=B_{\mbox{\tiny{flux}}}(Z)$ when Eq. (\ref{Z1})
  applied and $B_{\mbox{\tiny{sup}}}=B_{\mbox{\tiny{cut-off}}}(Z)$ when
  Eq. (\ref{Z2}) applied}
\end{figure*}


\subsection{Constraint deduced on $R$ and $Z$}
First of all, it seems likely that $R\geq R_s$, where $R_s$ is the
Schwarzschild radius of the black hole supposed to power the AGN. We
obtain a lower limit for $R_s$ through the Eddington limit. Assuming
$L_{\mbox{\tiny{UV}}}=4\pi D^2 F_{\mbox{\tiny{UV}}}$ as roughly equal to
the bolometric luminosity, it gives:
\begin{equation}
R\geq R_s\geq \frac{L_{\mbox{\tiny{UV}}}\sigma_t}{2\pi m_{p}c^3}.
\label{Rmin}
\end{equation}
On the contrary, the smaller X-ray time variability $\Delta t_{min}$ (if
known) gives an upper limit for the size of the non-thermal source:
\begin{equation}
R\leq R_{var}=c\Delta t_{min}.
\label{Rmax}
\end{equation}
Finally,  we must have at least:
\begin{equation}
 Z\geq R. 
\label{ZsupR}
\end{equation}
On the other hand, it appears from Eq.  (\ref{eqsyn}) that, to observe no
synchrotron emission at the I band frequency $\nu_{\mbox{\tiny{I}}}$, a
sufficient (but not necessary) condition is $\nu_{\mbox{\tiny{I}}} \geq
\nu_c$, that is the upper cut-off of the spectrum lies below our observed
frequency. It gives thus a possible upper limit for the strength of the
magnetic field:
\begin{equation}
B\leq B_{\mbox{\tiny{cut-off}}} \simeq
\frac{4\pi mc}{3q}\frac{\nu_{\mbox{\tiny{I}}}}{\gamma_{max}^2}=
\frac{C_{\mbox{\tiny{cut-off}}}}{\gamma_{max}^2}.
\label{Bcut}
\end{equation}
We can also constraint $\gamma_{max}$ since we know that the X-ray
spectrum of Seyfert galaxies can be fitted by a power law from $\simeq
1keV$ to $\simeq 100-500\ keV$, where an exponential cut-off is observed
(Jourdain et al. \cite{Jou92a}; Maisack et al. \cite{Mais93}; Dermer \&
Gehrels \cite{DermGehr95}). Since the mean frequency of the soft UV
photons is roughly in the range $5-50\ eV$ (Walter et al. \cite{Wal94}),
the maximum Lorentz factor $\gamma_{max}$ of the particles must be in the
range 50-300.\\ 
Besides, limits on $\sigma_{supp}$ resulting from our data analysis (see
Section \ref{dataanalysis}) give upper limits on the flux of the variable
component $F_{\mbox{\tiny{syn}}}=\nu_{\mbox{\tiny{I}}}F_{\nu}^{syn}$ for
each galaxy. Consequently, combining Eqs.(\ref{eqsyn}) and (\ref{eqIC})
we obtain another possible upper limit for the magnetic field:
\begin{eqnarray}
B\leq B_{\mbox{\tiny{flux}}} & = & \left(\frac{F(s)}{E(s)}\frac{D^2}{Z^2}
  \frac{F_{\mbox{\tiny{syn}}} 
  F_{\mbox{\tiny{UV}}}}{F_{\mbox{\tiny{X}}}}
  \left(\frac{\nu_{\mbox{\tiny{I}}}
  \nu_{\mbox{\tiny{UV}}}}{\nu_{\mbox{\tiny{X}}}}
  \right)^{\frac{s-3}{2}}\right)^{\frac{2}{s+1}}  \nonumber \\
 & = & C_{\mbox{\tiny{flux}}} Z^{-\frac{4}{s+1}}
\label{Bobs}
\end{eqnarray}
In this equation $\nu_{\mbox{\tiny{X}}}$ is the mean X-ray frequency
depending on the X-ray data for each objects, and
$F_{\mbox{\tiny{X}}}=\nu_{\mbox{\tiny{X}}}F_{\nu_{\mbox{\tiny{X}}}}^
{\mbox{\tiny{X}}}$ is the associated mean flux.  Thus, no
microvariability detection in any galaxy of our sample, means that:
\begin{equation}
 B\leq max(B_{\mbox{\tiny{cut-off}}},\ B_{\mbox{\tiny{flux}}})= B_{sup}.
 \label{Bmax}
\end{equation}
We have studied these differents constraints for only seven galaxies of
our sample whose UV and X-ray luminosity and spectral index are reported
in Walter \& Fink (\cite{Wal93}). These data are gathered together in
Table \ref{tab3}, with the corresponding values of $R_s$, $R_{var}$,
$F_{\mbox{\tiny{syn}}}$ and $B_{flux}$ for each of the galaxies. The
galaxy NGC 4051 is the only one for which a variability in the X-ray is
known, down to 100 s. As a conservative value to estimate the maximum
X-ray size for this galaxy, we use $\Delta t_{min}$ = 300 s.\\
Further constraint come from equipartition between particles and magnetic
field.  Effectively, non-thermal particles need to be accelerated to
compensate synchrotron and Inverse Compton losses and magnetic field is
generally invoked in the acceleration process (Fermi processes in a shock
for example). In this case the magnetic energy density must be equal or
larger than the particles energy density. Defining the equipartition
value $B_{eq}$ for the magnetic field:
\begin{equation}
  \frac{B^2_{eq}}{8\pi}=\frac{n_0 mc^2 \gamma_{min}^{2-s}}{s-2}
\end{equation}
and deducing $n_0$ from Eq.(\ref{eqIC}), we must have finally:
\begin{eqnarray}
  B\geq B_{eq} & = & \left(\frac{8\pi Z^2
      mc^2}{(s-2)\gamma_{min}^{2-s} R^3 F(s)} 
    \frac{F_{\tiny{X}}}{F_{\mbox{\tiny{UV}}}}
    \left(\frac{\nu_{\tiny{X}}}{\nu_{\mbox{\tiny{UV}}}}\right)^
    {\frac{s-3}{2}}\right)^{1/2} \nonumber \\
   & = & C_{eq}\frac{Z}{R^{3/2}}
  \label{Beq}
\end{eqnarray}
Inequalities (\ref{Bmax}) and (\ref{Beq}) reduce finally to inequalities
between $Z$ and $R$:
\begin{equation}
Z \leq \left( \frac{C_{flux}}{C_{eq}} R^{3/2} \right
)^{\frac{s+1}{s+5}} \hspace{0.5cm} (\text{corresponding to }
B_{eq}\leq  B_{flux}) \label{Z1}
\end{equation}
or
\begin{equation}
Z \leq \frac{C_{\mbox{\tiny{cut-off}}}}{C_{eq}}
\frac{R^{3/2}}{\gamma_{max}^2} \hspace{1cm} (\text{corresponding to }
B_{eq}\leq   B_{cut-off}) \label{Z2}
\end{equation}
Plots Z vs. R of Fig. \ref{ZvsR} compiled the constraints described
above. We have plotted the curves (type I) corresponding to constraint
(\ref{Z1}) for each galaxy in dashed line. The second inequality
(\ref{Z2}) gives a set of limiting curves (type II) on the assumed value
of $\gamma_{max}$. Since these curves represent the equipartition
$B_{eq}=B_{\mbox{\tiny{cut-off}}}(\gamma_{max})$, they can also be
considered as isocontours of $B_{eq}(R,Z)$. We have plotted type II
curves corresponding, from left to right, to $\gamma_{max}=50,\ 100,\ 
200,\ 300$, which correspond to $B_{eq}=32000\ G,\ 8000\ G,\ 2000\ G$ and
$1000\ G$. The diagrams must be read as follows:
\begin{enumerate}
\item For each galaxy, the allowed region is constrained by Eqs.
  (\ref{Rmin}), (\ref{Rmax}), (\ref{ZsupR}), (\ref{Z1}) and (\ref{Z2}).
  It is colored in grey in each plot for $\gamma_{max}=100$. Other
  $\gamma_{max}$ values would correspond to another curve of type II. The
  hashed regions are forbidden by Eqs. (\ref{Rmin}) and (\ref{Rmax}).
\item At a given point inside the allowed region, a lower limit of $B$ is
  given by $B_{\mbox{\tiny{eq}}}$, represented by the type II curve
  passing through this point. An upper limit is given by
  $B_{\mbox{\tiny{flux}}}$ if Eq.  (\ref{Z1}) applies or by
  $B_{\mbox{\tiny{cut-off}}}$ if Eq.  (\ref{Z2}) applies.
  $B_{\mbox{\tiny{flux}}}$, $B_{\mbox{\tiny{cut-off}}}$ and
  $B_{\mbox{\tiny{sup}}}$ are plotted on the right of each graphic.  The
  equality $B_{\mbox{\tiny{flux}}}=B_{\mbox{\tiny{cut-off}}}$ is
  realized, for a given assumed value of $\gamma_{max}$, when type I and
  type II curves intersect. An absolute maximum of the magnetic field is
  obtained for the smaller value of $Z$ in the allowed region.  This
  value is also reported in Table \ref{tab3}.
\end{enumerate}
An allowed region exists for each galaxy, with a critical case for NGC
4051, where the space parameter is strongly constrained. However our
results for this galaxy disagree with those of Celotti et al.
(\cite{Cel91}), since if we assume, like them, that the size of the X-ray
region is strictly equal to $R_{var}$, we are intside the allowed region
for non-thermal models. But these results need to be used with care, in
the case of this galaxy, since it seems unlikely for R and Z to be so
fine tuned.  These different results are obviously affected by the lack
of simultaneous X-ray and Optical-UV data and constraints could be
tightened if rapid X-ray variability were detected for most of these
objects. It appears however that non-thermal model can not be ruled out
by our data and can still explain the high energy spectra of Seyfert
galaxies.

\section{Conclusion}
Upper limits on optical microvariabilities in a large sample of 22
Seyfert galaxies have been obtained, using differential photometry. We
have developped a new method of analysis minimizing the influence of
possible variability of the comparison stars. We thus obtain precision on
our variability detection smaller than 1\% and in most cases about 0.5\%.
We do not detect variability in any of our objects, with a possible trend
of several hours in Mkn 359. In the hypothesis where variable optical
emission would be due to synchrotron radiation from the non-thermal
electron population which we suppose to be responsible for the X-ray
emission, these results enable us to constraint intrinsic properties of
the local environment of the non thermal source. Upper limits on our
variability detection and equipartition hypothesis between magnetic field
and particle, restrain the possible values of the size R of the non
thermal source, its distance Z from the UV emission region and fix upper
and lower limits for the magnetic field.\\

\noindent
{\sl Acknowlegements:} We are grateful to the referee, Prof. Gopal
Krishna, for his precious comments and really careful reading of the
manuscript. We feel that his suggestions and criticisms have certainly
improved this paper.

\setcounter{figure}{6}
\begin{figure*}[tbp]
\begin{tabular*}{\textwidth}{@{\excs}ll@{\extracolsep{0pt}}}
\includegraphics[width=\columnwidth]{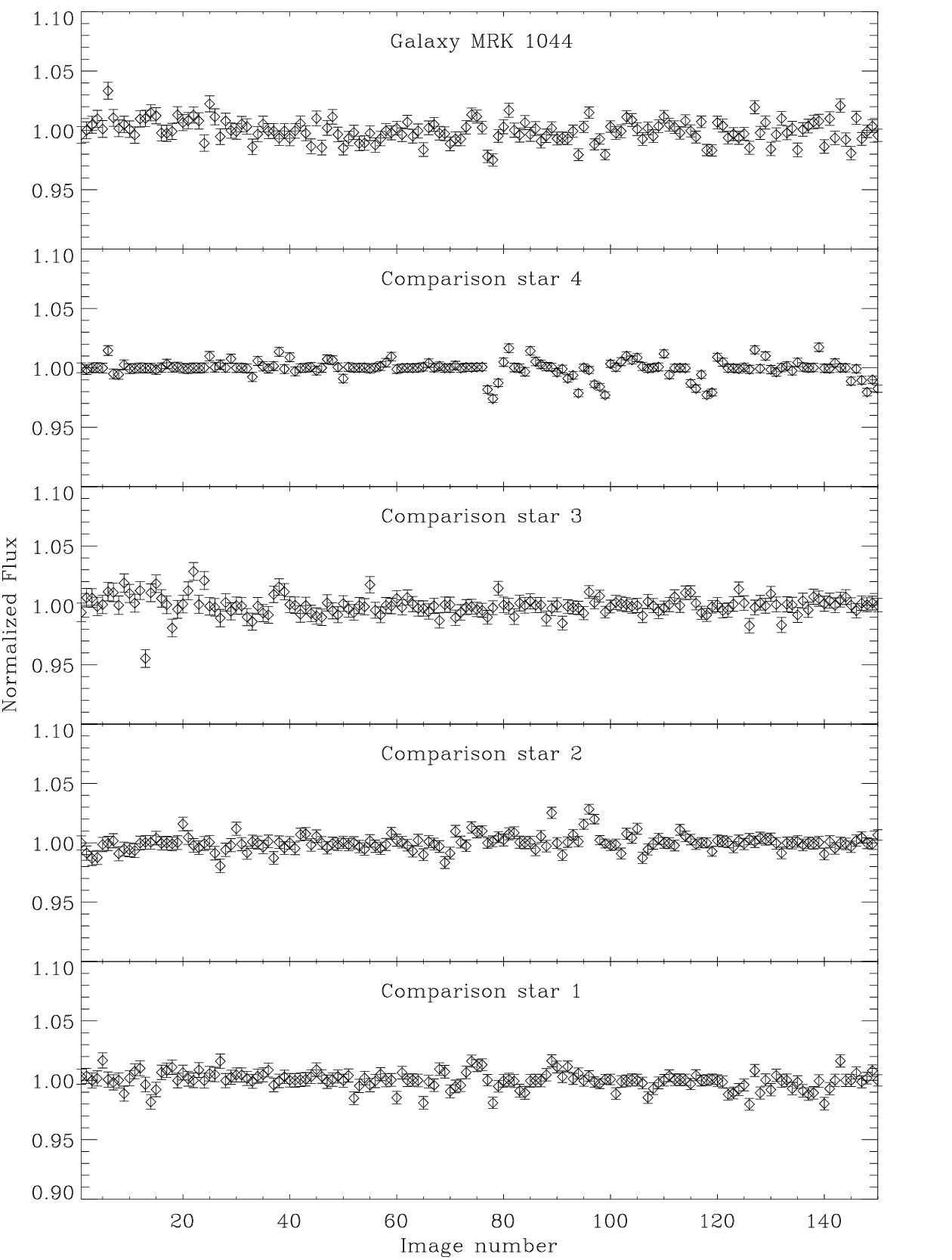} & 
\includegraphics[width=\columnwidth]{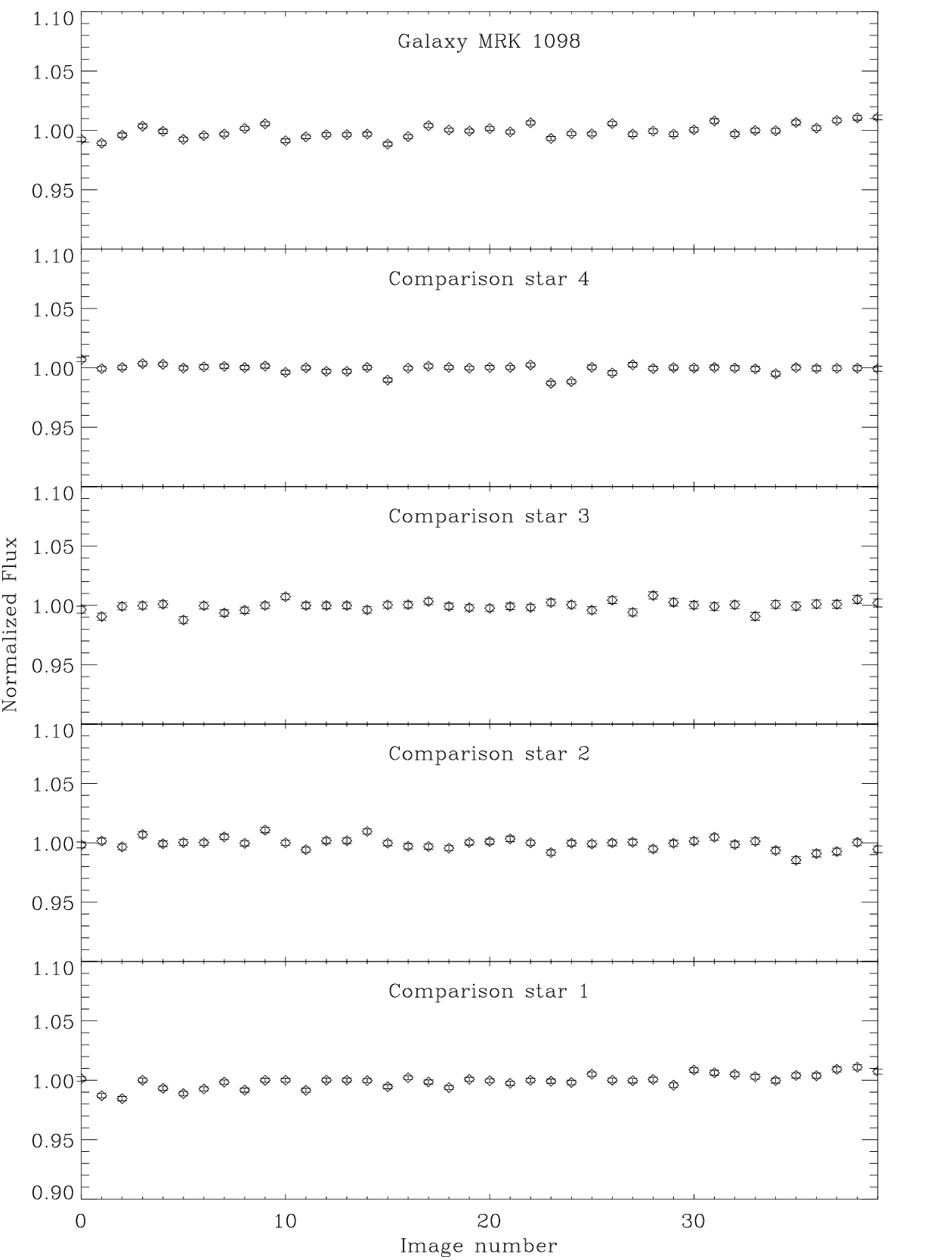} \\
\includegraphics[width=\columnwidth]{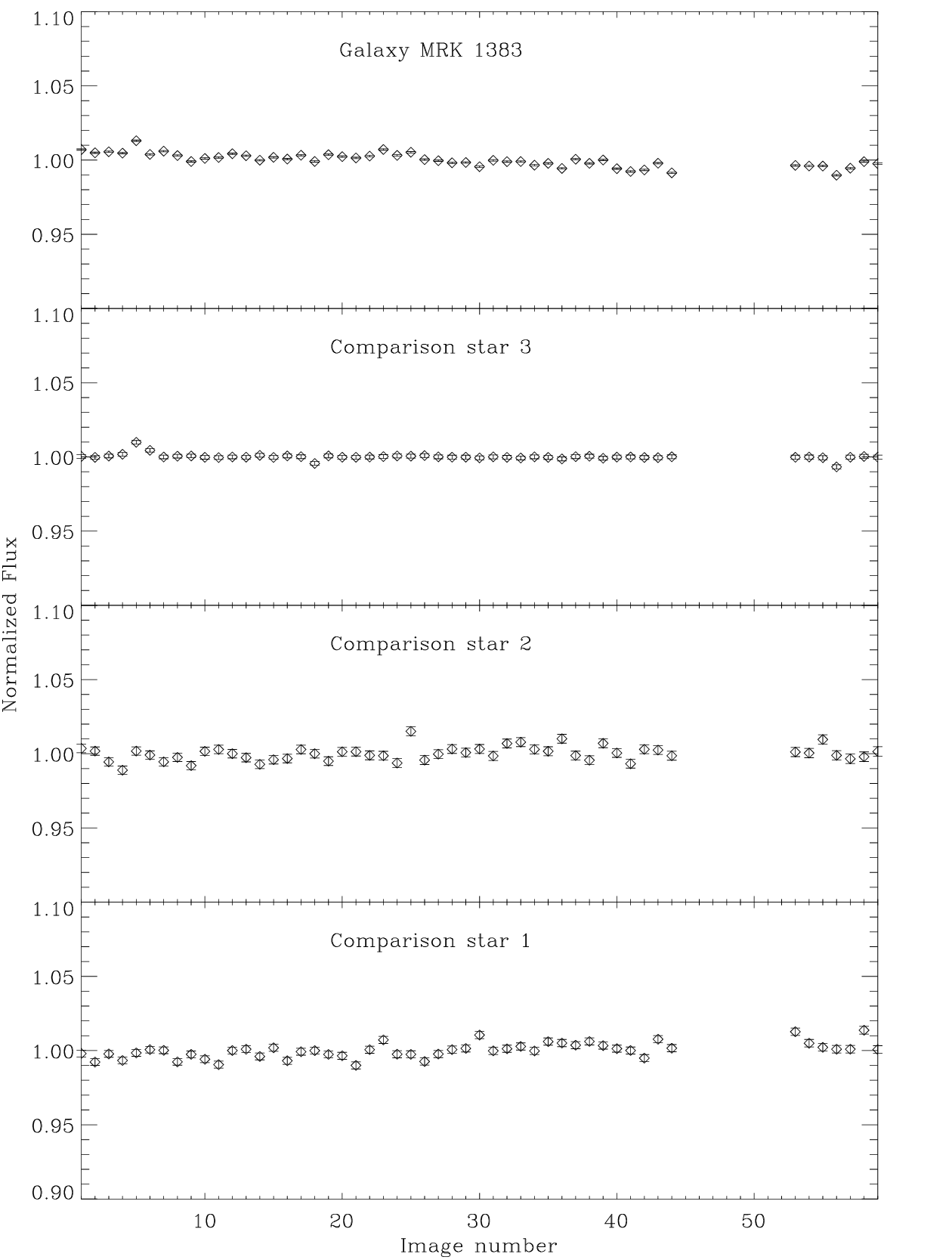} &
\includegraphics[width=\columnwidth]{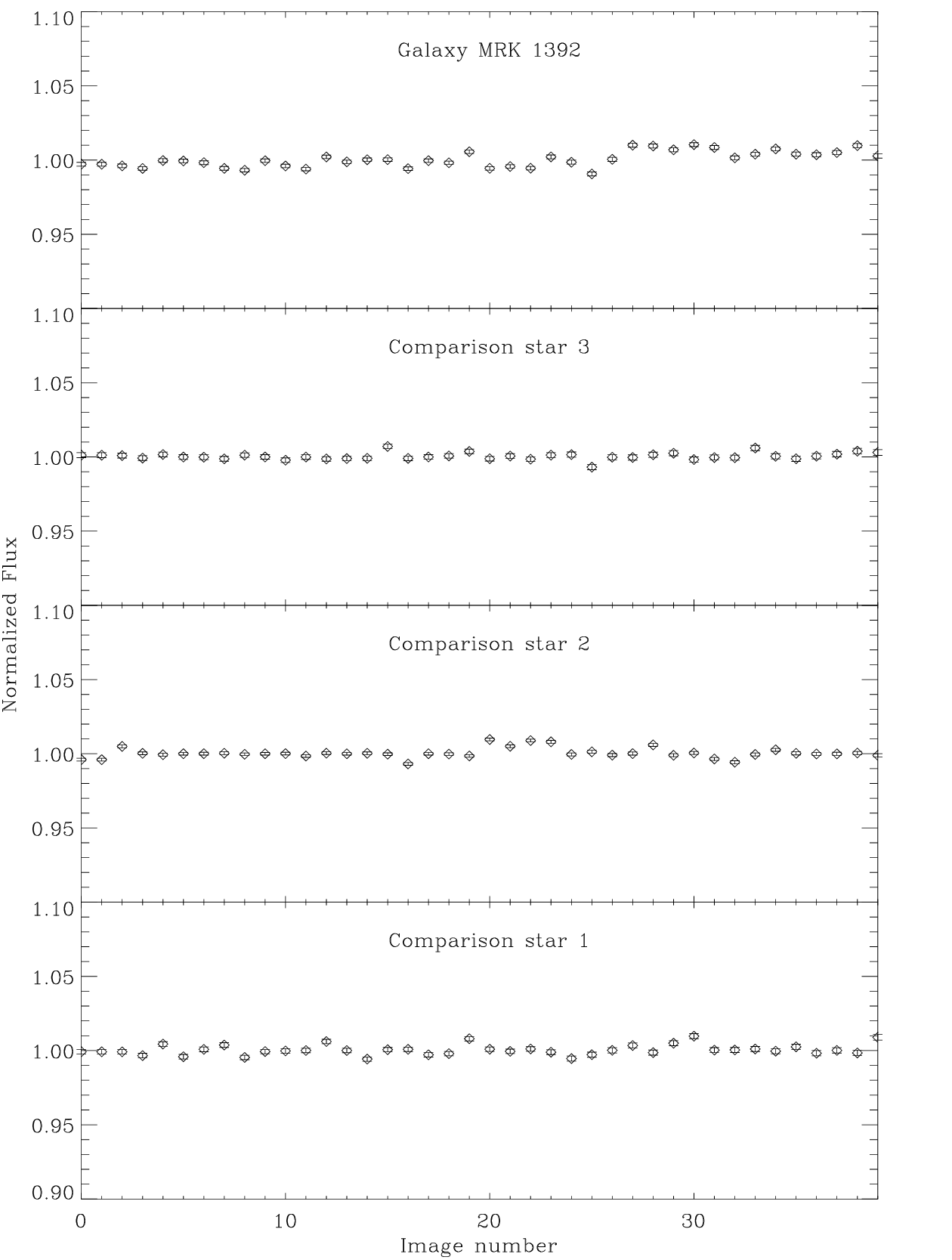} \\
\end{tabular*}
\caption[]{To be continued.} 
\end{figure*}

\setcounter{figure}{6}
\begin{figure*}[tbp]
\begin{tabular*}{\textwidth}{@{\excs}ll@{\extracolsep{0pt}}}
\includegraphics[width=\columnwidth]{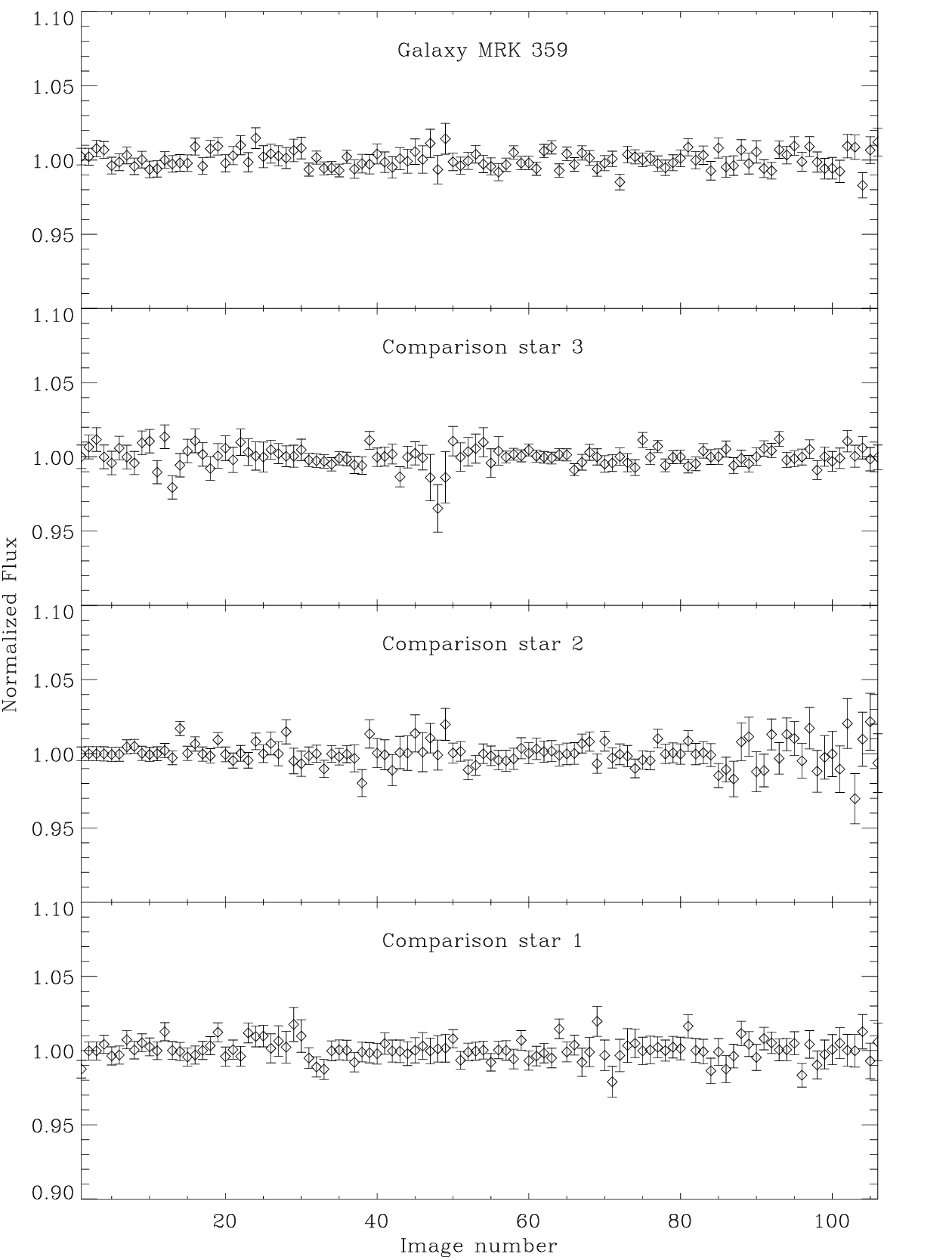} & 
\includegraphics[width=\columnwidth]{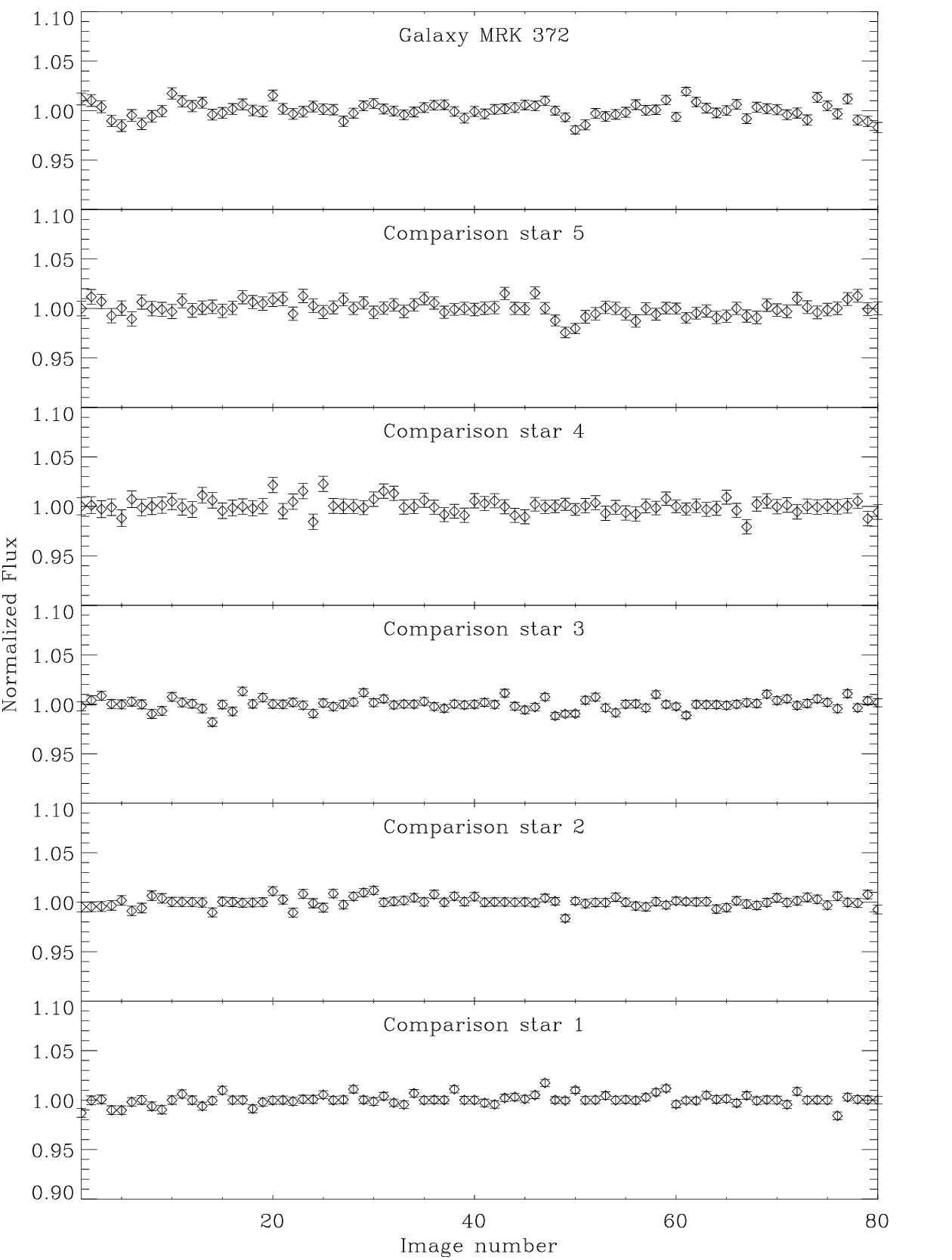} \\
\includegraphics[width=\columnwidth]{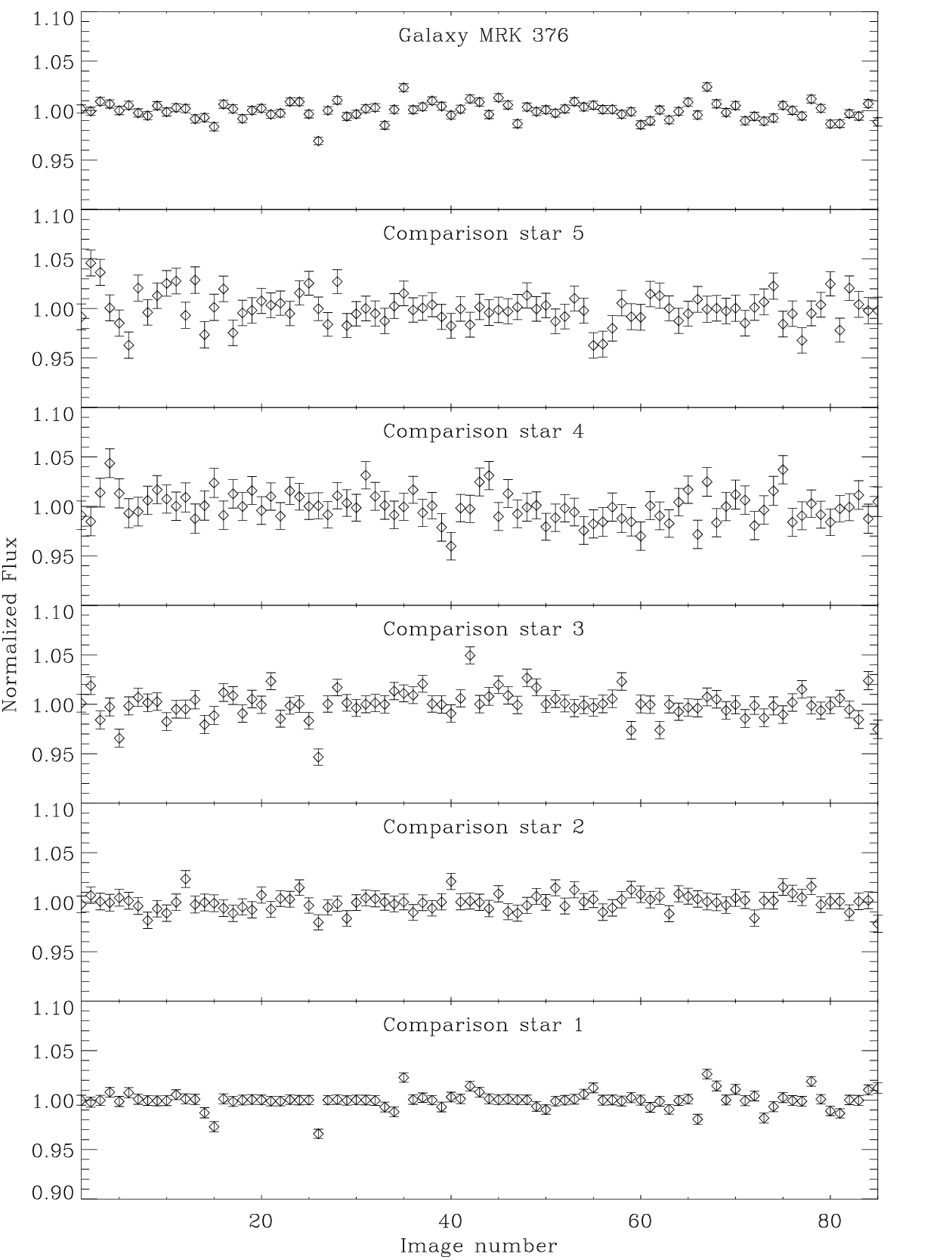} &
\includegraphics[width=\columnwidth]{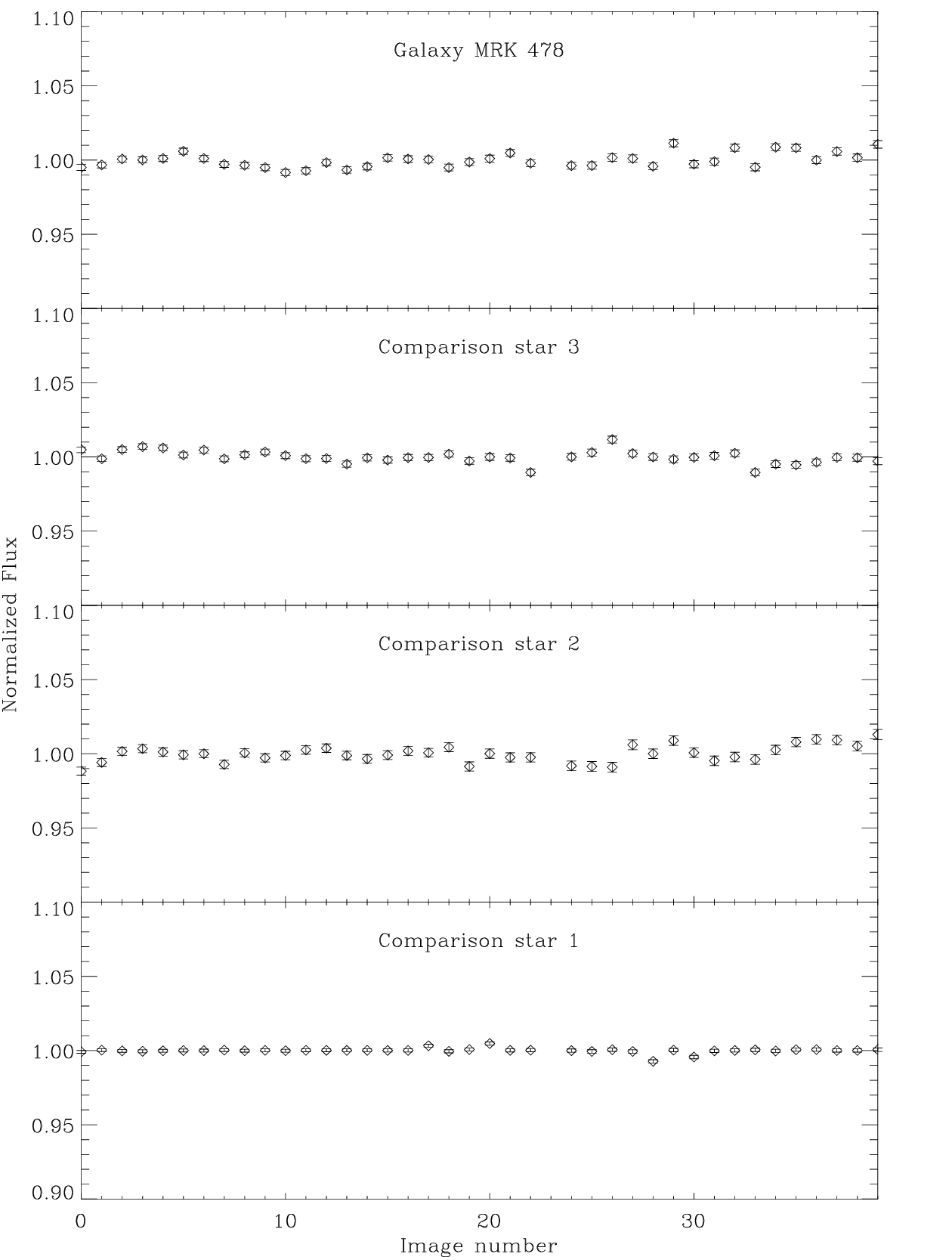} \\
\end{tabular*}
\caption[]{To be continued.} 
\end{figure*}

\setcounter{figure}{6}
\begin{figure*}[tbp]
\begin{tabular*}{\textwidth}{@{\excs}ll@{\extracolsep{0pt}}}
\includegraphics[width=\columnwidth]{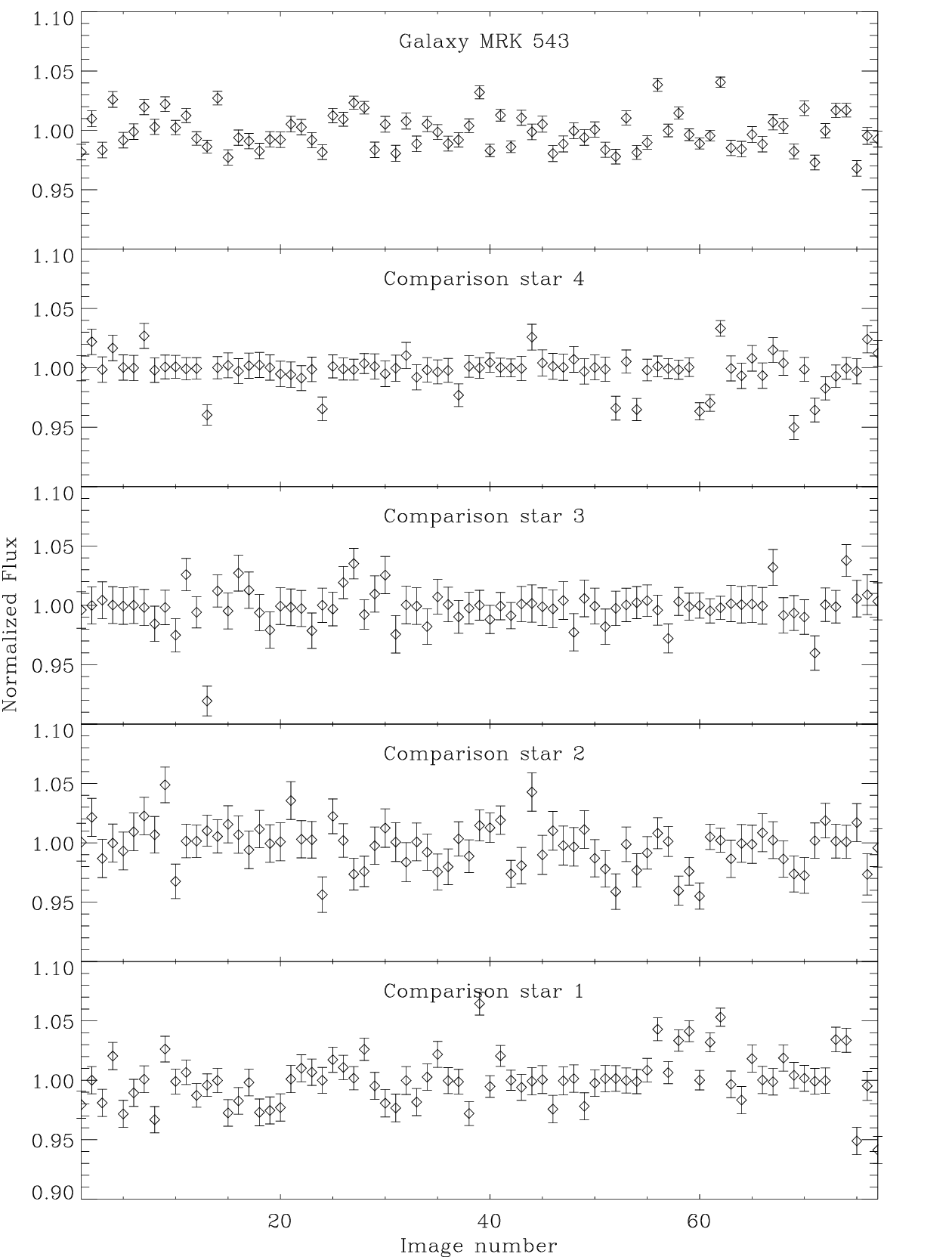} & 
\includegraphics[width=\columnwidth]{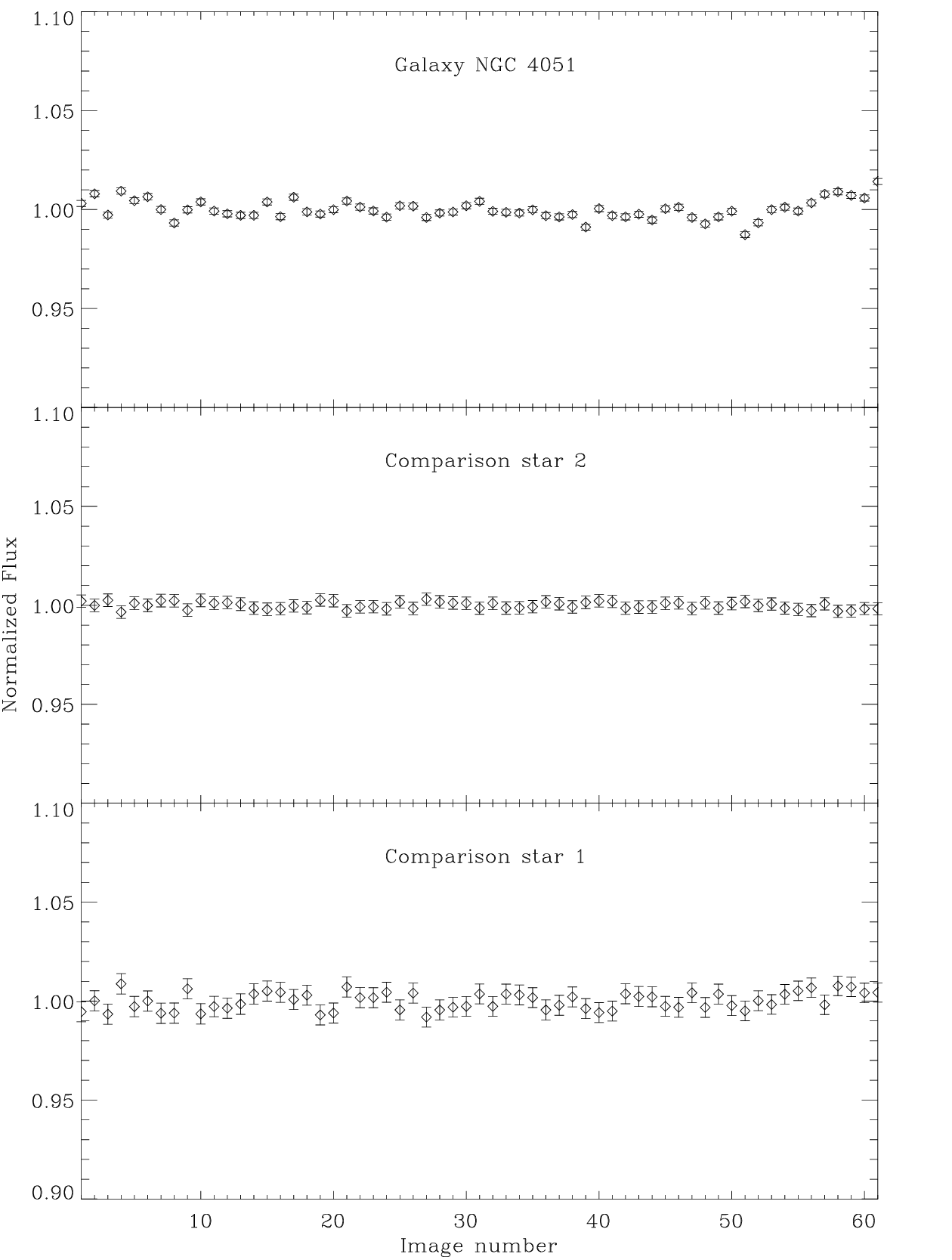} \\
\includegraphics[width=\columnwidth]{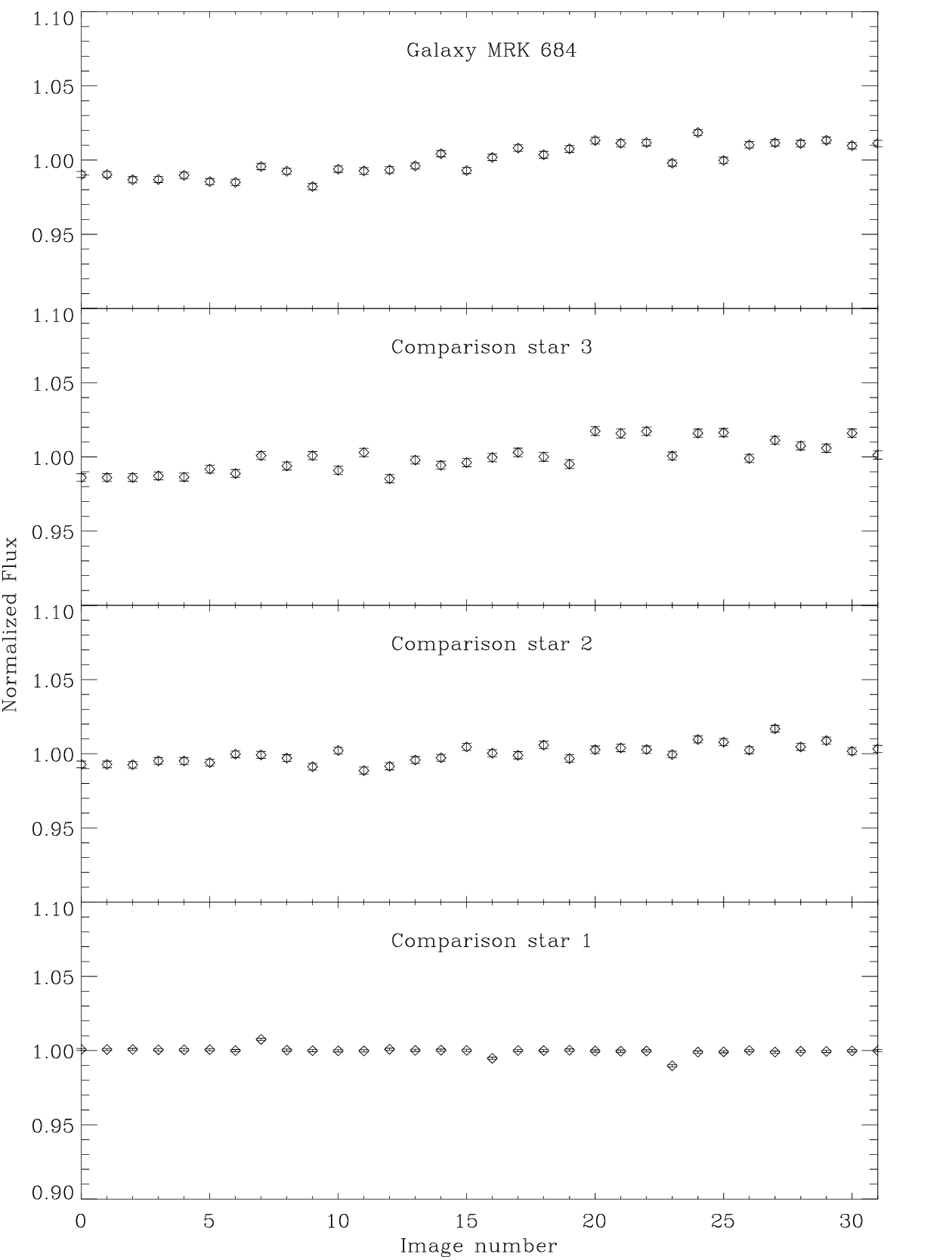} &
\includegraphics[width=\columnwidth]{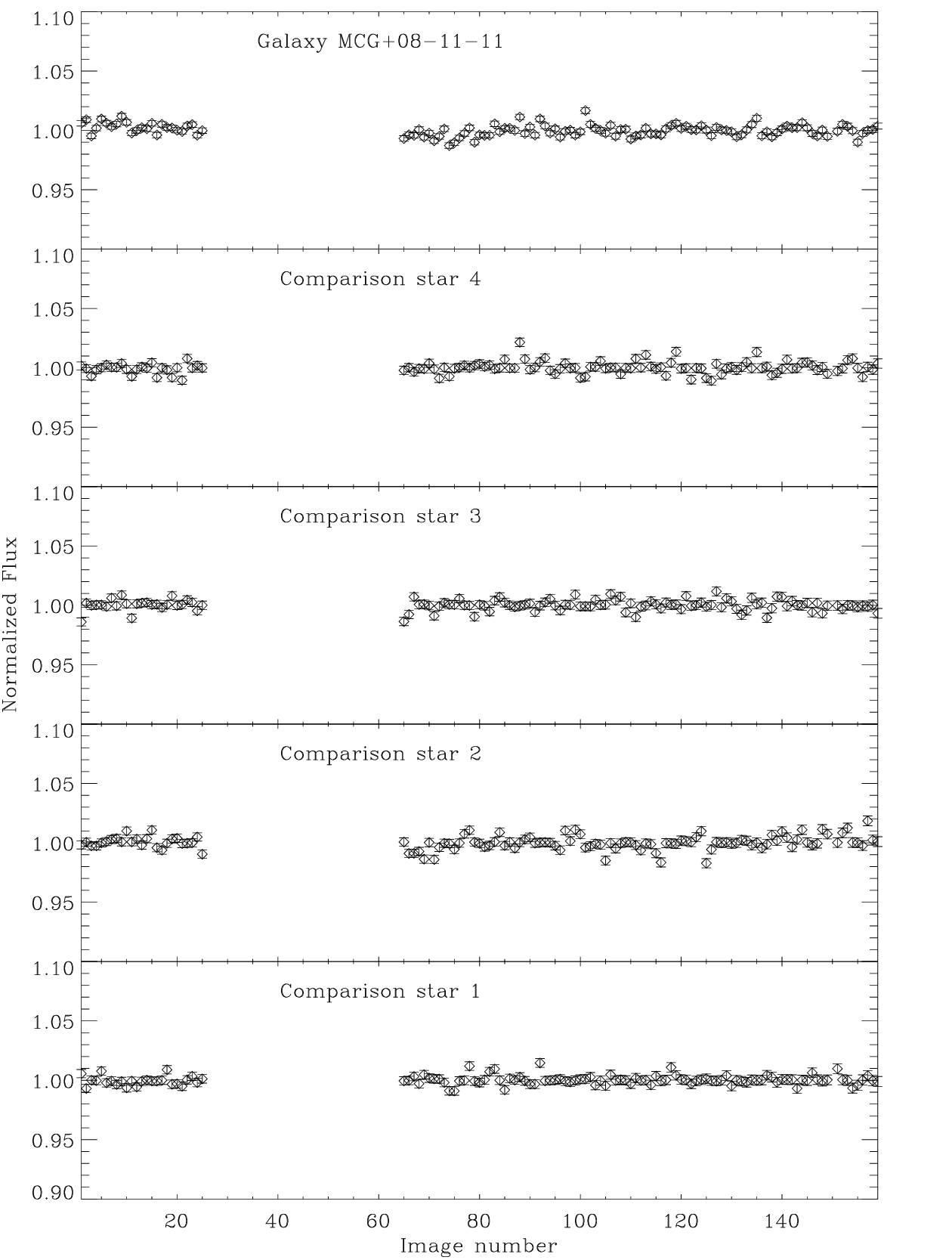} \\
\end{tabular*}
\caption[]{To be continued.} 
\end{figure*}

\setcounter{figure}{6}
\begin{figure*}[tbp]
\begin{tabular*}{\textwidth}{@{\excs}ll@{\extracolsep{0pt}}}
\includegraphics[width=\columnwidth]{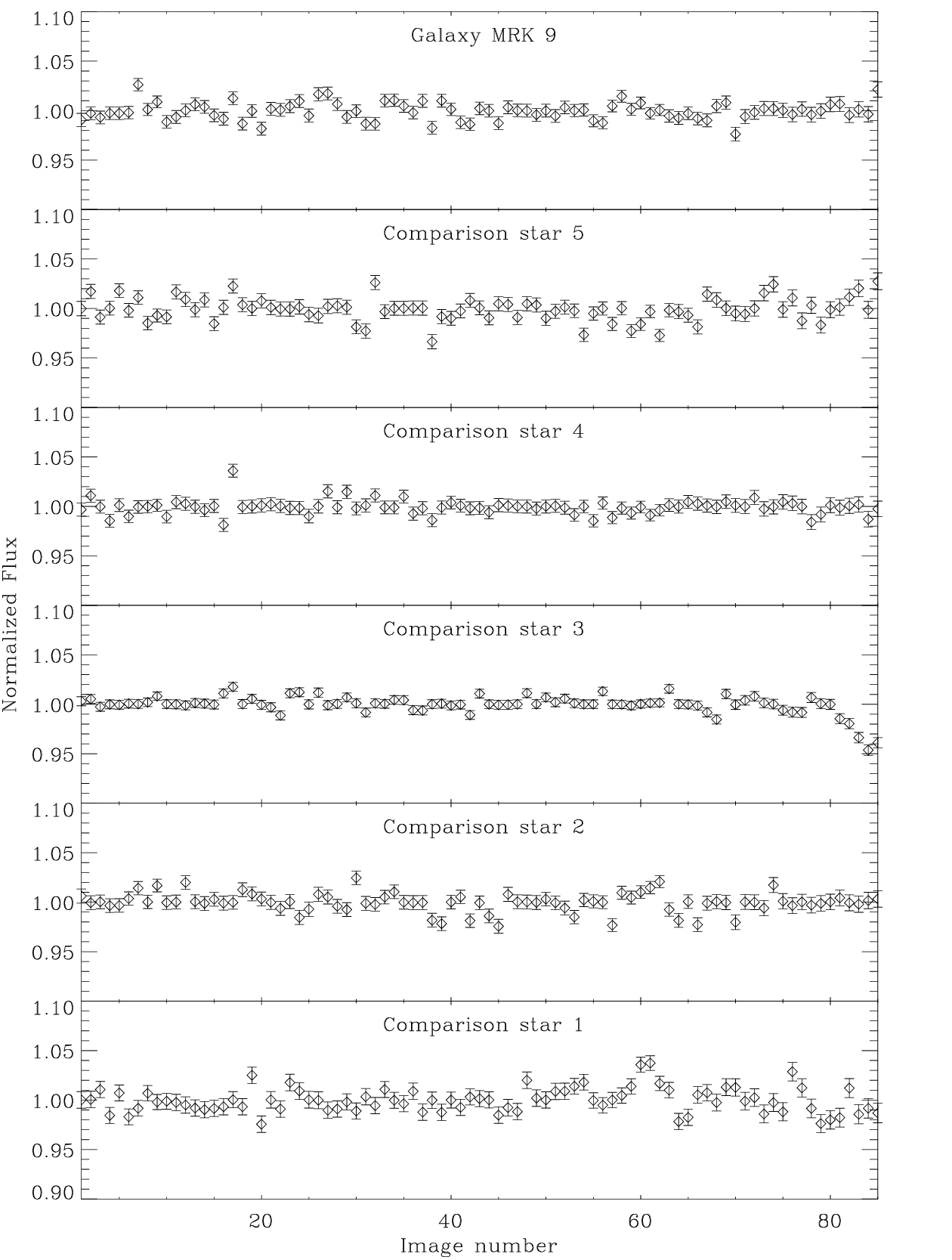} & 
\includegraphics[width=\columnwidth]{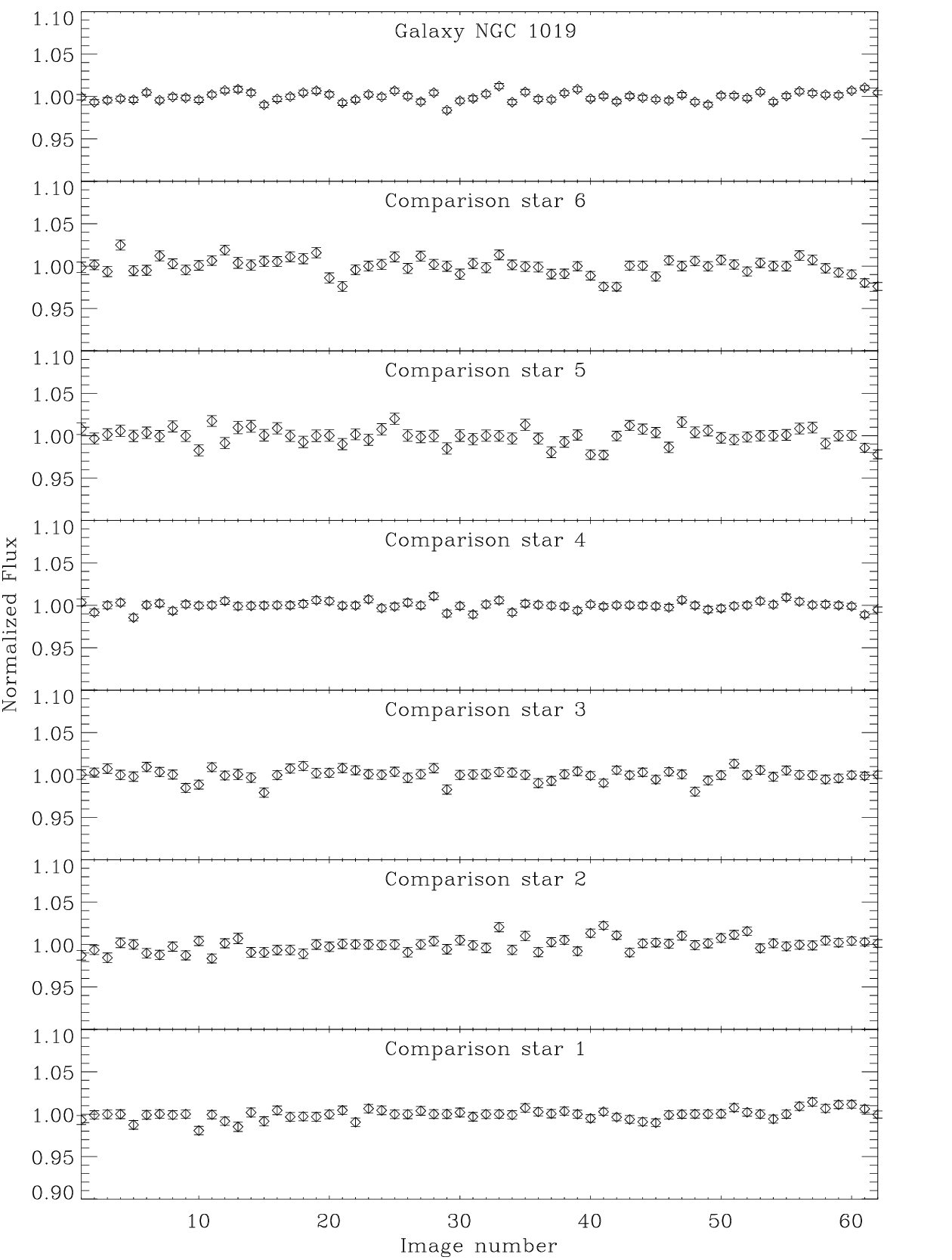} \\
\includegraphics[width=\columnwidth]{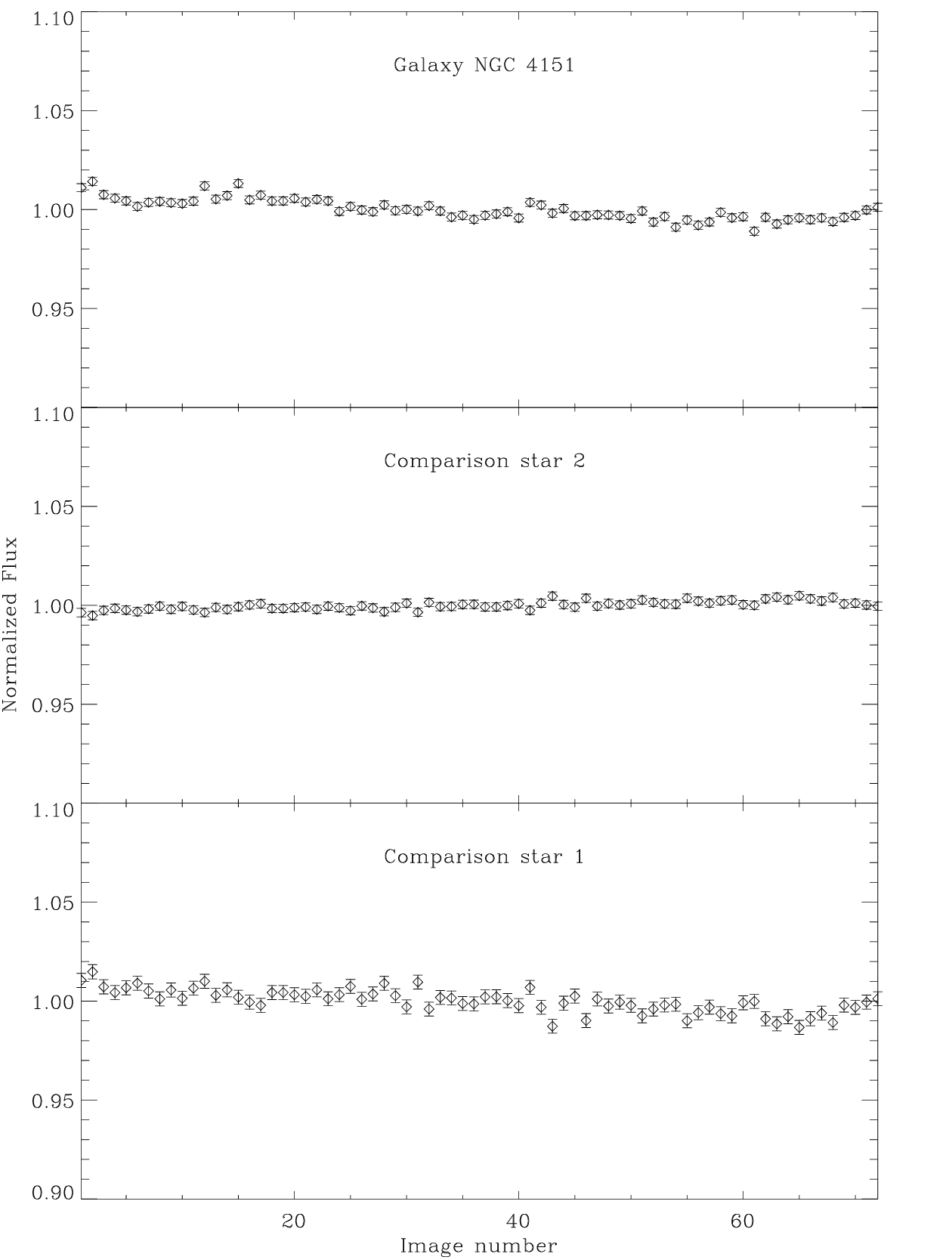} &
\includegraphics[width=\columnwidth]{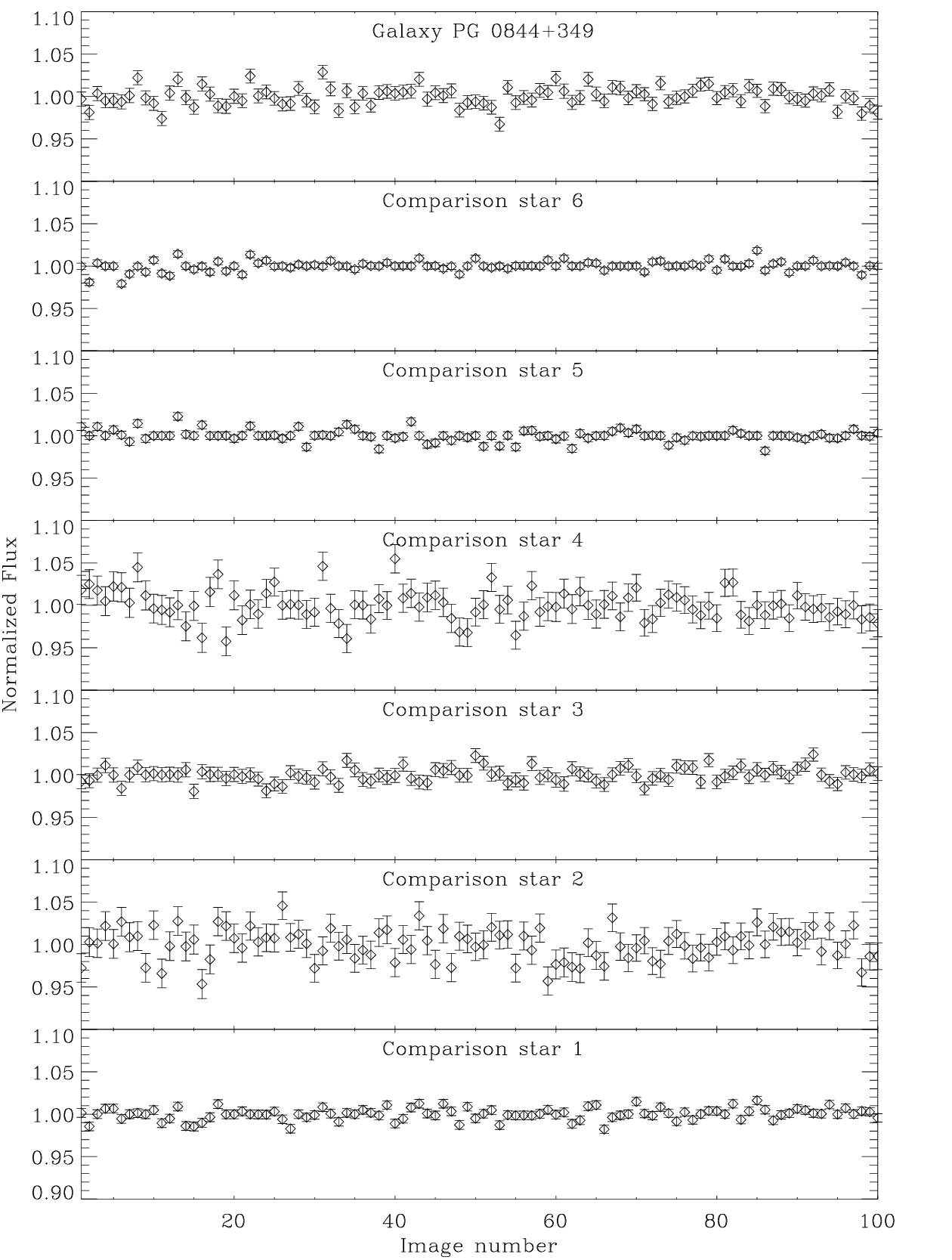} \\
\end{tabular*}
\caption[]{To be continued.} 
\end{figure*}

\setcounter{figure}{6}
\begin{figure*}[tbp]
\begin{tabular*}{\textwidth}{@{\excs}ll@{\extracolsep{0pt}}}
\includegraphics[width=\columnwidth]{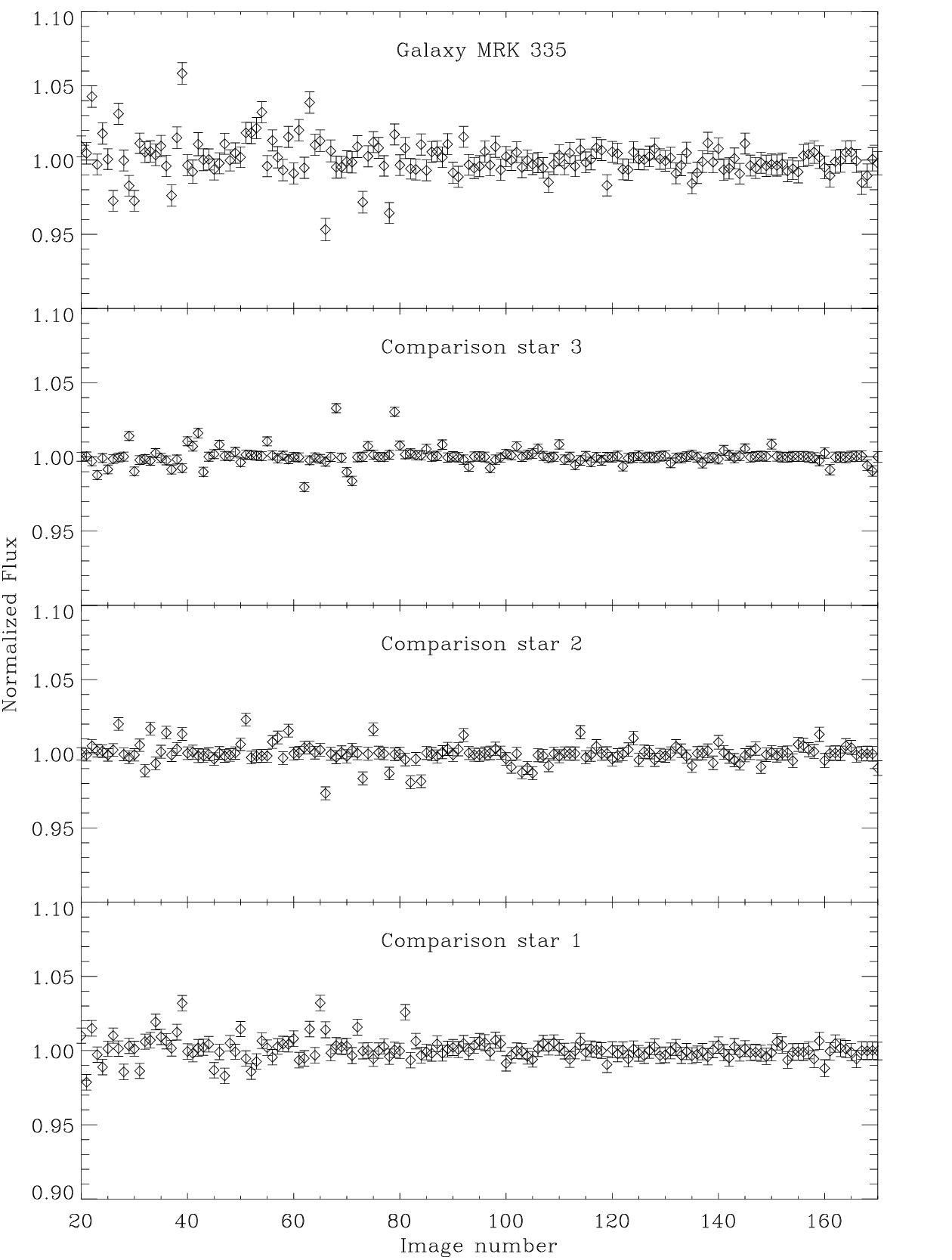} \\ 
\end{tabular*}
\caption[]{To be continued.} 
\end{figure*}

\begin{thebibliography}{}
  
\bibitem[1989]{Asl89}
Aslanov A. A., Kolosov, D. E., Lipunova, N. A., et al., 1989, SvAL 15, 132

\bibitem[1983]{Bar83}
Barr P., Willis A.J., Wilson R., 1983, MNRAS 203, 201

\bibitem[1970]{Blu70}
Blumenthal G. R., Gould R. J., 1970, Rev. Mod. Phys. 42, 237

\bibitem[1990]{Car90}
Carini, M. T. 1990, The time scales of the optical variability of
blazars, Ph.D. Thesis, 11  

\bibitem[1991]{Car91}
Carini, M. T., et al., 1991, AJ 101, 1196
    
\bibitem[1992]{Car92}
Carini, M. T., Miller, H. R., Noble, J. C., et al., 1992, AJ 104, 15

\bibitem[1991]{Cel91}
Celotti, A., Ghisellini, G., Fabian A. C., 1991, MNRAS 251, 529

\bibitem[1993]{Cha93}
Chakrabarti S. K. \& Wiita P. J., 1993, ApJ 411, 602

\bibitem[1996]{Col96}
Colbert, E. J. M., Baum, S. A., Gallimore, J. F., O'Dea, C. P., \&
Christensen, J. A. 1996, ApJ 467, 551

\bibitem[1993]{Cru93}
Cruz-Gonz\`{a}lez, I., Carrasco, L., Ruiz, E., et al., 1993,
Rev. Mex. Astron. Astrof. 29, 197

\bibitem[1995]{DermGehr95}
Dermer C.D., Gehrels N., 1995, ApJ 447, 103

\bibitem[1990]{Don90}
Done C., Ward M. J., Fabian A. C., et al., 1990, MNRAS 243, 713 

\bibitem[1992]{Dor92}
Doroshenko, V. T., Lyuty\u{i} V. M., Sillanp\"{a}\"{a}, A., Valtaoja, E.,
1992, in Variability of Blazars, eds. E. Valtaoja \& M. Valtonen
(Cambridge: University Press), p. 358  

\bibitem[1992]{Dul92}
Dultzin-Hacyan, D., Schuster, W., Parrao, L., et al., 1992, AJ 103, 1769

\bibitem[1993]{Dul93}
Dultzin-Hacyan,, Ruelas-Mayorga, A., Costero, R., 1993,
Rev. Mex. Astron. Astrof. 25, 143

\bibitem[1965]{Gin65}
Ginzburg V. L. and Syrovatskii S. I., 1965, ARAA 3, 297 

\bibitem[1995]{Gop95}
Gopal-Krishna, Ram Sagar, Wiita P. J., 1995, MNRAS 274, 701

\bibitem[1993]{Gop93}
Gopal-Krishna, Wiita P. J., 1992, A\&A 259,109

\bibitem[1992]{Gop92}
Gopal-Krishna,  Ram Sagar, Wiita P. J., 1993, MNRAS 262, 969

\bibitem[1992]{Gra92}
Grandi, P., Tagliaferri, G., Giommi, P., Barr, P., \& Palumbo,
G. G. C. 1992, ApJS, 82, 93  

\bibitem[1991]{Haa91}
Haardt F. \& Maraschi L., 1991, ApJ 380, 51

\bibitem[1997]{Hen97}
Henri G. \& Petrucci P. O., 1997, A\&A 326, 87

\bibitem[1992]{Hun92}
 Hunt, L. K., Mannucci, F., Salvati, M., \& Stanga, R. M. 1992, A\&A, 257, 
434 

\bibitem[1997]{Jan95}
Jang M. \& Miller H. R., 1995, ApJ 452, 582

\bibitem[1997]{Jan97}
Jang M. \& Miller H. R., 1997, AJ 114 (2), 565

\bibitem[1992]{Jou92a}
Jourdain E., et al., 1992a, A\&A 256, L38

\bibitem[1997]{Koe97}
Koenig M., Staubert R., Wilms J., 1997, A\&A 326, L25

\bibitem[1992]{Lan92}
Landolt A. U., 1992, AJ 104, 340


\bibitem[1993]{Lyu93}
Lyuty\u{i} V. M., Doroshenko, V. T., 1993, SvAL 19, 405

\bibitem[1989]{Lyu89}
Lyuty\u{i} V. M., Aslanov A. A., Khruuzina T. S., et al., 1989, SvAL 15, 247

\bibitem[1981]{Law81}
Lawrence A., Giles A. B., Mc Hardy I. M. and Cooke B. A., 1981, MNRAS
195, 149 

\bibitem[1994]{Mal94}
Malaguti G., Bassani L., Caroli E., 1994, ApJS 94, 517

\bibitem[1985]{McH85}
McHardy I., 1985, SSRv 40, 559 

\bibitem[1993]{Mais93}
Maisack M., et al., 1993, ApJ 407, L61

\bibitem[1993]{Man93}
Mangalam A. V., Wiita P. J., 1993, ApJ 406,420

\bibitem[1992]{Mil92}
Miller H. R., Carini M. T., Noble J. C., Webb J. C., Wiita P. J., 1992,
in Variability of Blazars, eds. E. Valtaoja \& M. Valtonen 
(Cambridge: University Press), p. 320

\bibitem[1993]{Mus93}
Mushotzky, R. F., Done, C., \& Pounds, K. A. 1993,, ARAA 31, 717 

\bibitem[1997]{Pal97}
Paltani S. et al., 1997, A\&A 327, 539

\bibitem[1970]{Pac70}
Pacholczyk, A. G. 1970, Radio Astrophysics (W. H. Freeman: San
Francisco) 

\bibitem[1991]{Qia91}
Qian S. J., Quirrenbach A., Witzel A., et al., 1991, A\&A 241, 15

\bibitem[1989]{Ver89}
V\'eron-Cetty M.-P., V\'eron P., 1989, ESO Sci. Rep. No. 13

\bibitem[1998]{Rab98}
Rabette 1998, A\&AS, in press

\bibitem[1978]{Rut78}
Rutman J., 1978, Proc. IEEE 66, 1048

\bibitem[1989]{Ryb89}
Rybicki, G., B. \& Lightman, A., P., 1979, in Radiative processes in
Astrophysics, Wiley-interscience, New-York.

\bibitem[1985]{sim85}
Simonetti J.H., Cordes J.M. \& Heeschen D.S., 1985, ApJ 296, 46

\bibitem[1997]{ulr97}
Ulrich, M. H., Maraschi, L., Megan, C., 1997, ARA\&A 35, 445 

\bibitem[1992]{Wag92}
Wagner S., 1992, in Variability of Blazars, eds. E. Valtaoja \& M. Valtonen
(Cambridge: University Press), p. 346  

\bibitem[1992]{Wal92}
Walter R. \& Courvoisier T. J.-L., 1992, A\&A 266, 65-71

\bibitem[1992]{Wal93}
Walter R. \& Fink H. H., 1993, A\&A 274, 105 

\bibitem[1994]{Wal94}
Walter et al., 1994, A\&A 285, 119

\bibitem[1991]{Wii91}
Wiita P. J., Miller H. R., Carini M. T., Rosen A., 1991, in structure and
Emission Properties of Accretion, IAU Colloquium No. 129, edited by
J. P. Lasota et al. (Editions Frontieres, Gif sur Yvette), p. 557

\bibitem[1992]{Wii92}
Wiita P. J., Miller H. R., Gupta N., Chakrabarti S. K., 1992, in
Variability of Blazars, eds. E. Valtaoja \& M. Valtonen 
(Cambridge: University Press), p. 311

\bibitem[1993]{Wil93}
Wilson, A. S. 1993, in Astrophysical Jets, ed. D. Burgarella, M. Livio \&
C.P. O'Dea (Cambridge: Cambridge Univ. Press), 121

\bibitem[1994]{Zdz94}
Zdziarski, A. A., Fabian, A. C., Nandra, K., 1994  MNRAS 269, L55 

\end{thebibliography}
\end{document}